\documentclass[preprint]{revtex4-1}
\usepackage{graphicx,subfigure}
\usepackage{amsfonts}
\usepackage{epsfig}
\usepackage[usenames]{color}

\begin{document}

\title{Nonlinear waves in second order conformal hydrodynamics}

\author{D.~A.~Foga\c{c}a, H.~Marrochio, F.~S.~Navarra, and J.~Noronha}
\address{Instituto de F\'{\i}sica, Universidade de S\~{a}o Paulo\\
C.P. 66318, 05315-970 S\~{a}o Paulo, SP, Brazil}

\vspace{2cm}

\begin{abstract}

In this work we study wave propagation in dissipative relativistic fluids described by a simplified set of the 2nd order viscous conformal hydrodynamic equations corresponding to Israel-Stewart theory. Small amplitude waves are studied within the linearization approximation while waves with large amplitude are investigated using the
reductive perturbation method, which is generalized to the case of 2nd order relativistic hydrodynamics. Our results indicate the presence of a ``soliton-like'' wave solution in Israel-Stewart hydrodynamics despite the presence of dissipation and relaxation effects.

\end{abstract}

%\pacs{PACS Numbers :  25.75.-q  24.10.Nz, 03.65.Pmjn  }
\maketitle

\vspace{1cm}

\section{Introduction}

Experiments at RHIC and LHC indicate that the quark-gluon plasma is an almost perfect fluid where viscous effects are small
\cite{Gyulassy:2004zy,new-qgp,Heinz:2013th}. In contrast, at low temperatures and nonzero baryon chemical potentials in the hadron gas phase, viscous effects may be considerably more pronounced \cite{NoronhaHostler:2008ju,Demir:2008tr,NoronhaHostler:2012ug,viscc,Denicol:2013nua}.
Numerical studies of the hydrodynamical evolution of the Quark-Gluon Plasma (QGP) show that viscosity produces some visible
but not very large effects on global observables \cite{visc-eff}.

In this work we investigate how the presence of a nonzero shear viscosity relaxation time affects wave propagation in relativistic fluids.
Waves in a hadronic medium may be caused, for example, by fluctuations in
baryon number or energy density. These fluctuations may be produced by inhomogeneous
initial conditions which, as pointed out in \cite{kapu}, are the result of quantum fluctuations in the
densities of the two colliding nuclei and also in the energy deposition mechanism. These
fluctuations and their phenomenological implications have been studied extensively \cite{iniflu,peter,shuryak,flor,Andrade:2013poa}
because they may be responsible for the angular correlations of particle emission observed in
heavy-ion experiments. There are also hydrodynamic fluctuations \cite{kapu}, which are the result of
finite particle number effects in a given fluid cell. This generates local thermal fluctuations of
the energy density (and flow velocity) which propagate throughout the fluid. Furthermore,
there may be fluctuations induced by energetic partons, which have been scattered in the
initial collision of the two nuclei and propagate through the medium, losing energy and
acting as a source term for the hydrodynamical equations \cite{jets}. Finally, there may be also freeze-out
fluctuations, which may be caused by finite particle number effects during and after the
freeze-out of the hydrodynamically expanding fluid.

In non-relativistic fluid dynamics the most successful theory of dissipative systems is the Navier-Stokes (NS) theory \cite{land,wein}.
For instance, one can use this theory to investigate the evolution of density perturbations in a non-relativistic hadron gas and in
a non-relativistic quark gluon plasma. Perturbations are usually studied with the linearization formalism \cite{roma,hidro},
which is the simplest way to study small deviations from equilibrium to obtain
wave equations, eventually featuring dissipative and relaxation terms. The propagation of perturbations through a QGP has been investigated
in several works with the help of a linearized version of the hydrodynamics of perfect fluids and of viscous fluids.
In \cite{fgn13}  the authors went  beyond linearization and considered
the effects of shear viscosity on the propagation of nonlinear waves. This study was performed with the help of the
well established reductive perturbation method \cite{davidson,rpm,leblond,loke}.

The simplest extension of the well-known Navier-Stokes equations to relativistic fluids is plagued with
instabilities and acausal signal propagation in the resulting equations \cite{acausal,shipu1,shipu2},
and, thus, they are not usually employed in numerical simulations.
Currently, most fluid-dynamical simulations of the QGP employ a set of relaxation-type equations similar to those derived by Israel and
Stewart (IS) \cite{ist} to close the conservation laws.

In this work we study the propagation of linear and nonlinear waves in relativistic fluids described by (a simplified set of) of the 2nd order conformal IS equations.
We show how to obtain the linear wave equation which contains the dissipative and relaxation terms.  This wave equation provides a dispersion relation that allows for the study of the stability and causality properties of the theory. We then expand the modes of the dispersion relation in powers of the wavenumber to investigate the effects of a nonzero relaxation time coefficient.  Understanding the relaxation effects in the linear modes proved to be a useful guide to approach the same question in the case of nonlinear perturbations.

Solutions of nonlinear equations of motion generally contain nonlinear dispersive and dissipative terms.  The relative
strength of these different terms depends on microscopic properties of the system, which manifest themselves in the transport coefficients (such as the shear viscosity coefficient, $\eta$, and the relaxation time, $\tau_{\pi}$) and in the equation of state. Different combinations of these terms generate Korteweg - de Vries solitons, shock waves, strongly damped waves and so on. In principle, given the underlying microscopic theory, one can calculate the transport coefficients and the equation of state to determine the type of waves which can propagate in the system. However, when the underlying theory is QCD this is not an easy task due to the strongly coupled nature of the quark-gluon plasma and one needs to resort to phenomenological models to estimate $\eta$ and $\tau_{\pi}$ (the QCD equation of state can be reliably computed on the lattice \cite{lattice}). The choice of these quantities defines the properties of the solutions of the wave equations. Inconsistent choices may lead to unphysical solutions and this opens the possibility of using waves to put some additional constraints on the values of  $\eta$ and $\tau_{\pi}$ in the QGP.

In this paper we show how to obtain a system of two coupled differential equations to study nonlinear waves in conformal IS theory. We solve this system numerically, determining the role played by shear viscosity and its relaxation time on wave packet evolution. One of the equations of this system is the Burgers' equation for the first order perturbation in the energy density, which does not contain relaxation effects, and it is, thus, the same equation obtained in the NS-based approach developed in \cite{nos2012}.  The other equation describes the second order perturbation in the energy density, where the effects from a nonzero relaxation time coefficient become manifest. Our results indicate the presence of a ``soliton-like'', i.e., an approximately solitary wave solution, in IS theory despite the dissipative and relaxation effects. Another motivation to
study nonlinear perturbations is the possibility to find an upper bound for the relaxation scale, which is not found in the linear treatment.

This paper is organized as follows. In the next section we review the basic expressions of the simplified set of equations of Israel-Stewart theory used in this work. In Section III we study the linearized hydrodynamic equations, derive the corresponding wave equation, and perform a study of the stability and
causality regarding the propagation of these waves. In Section IV we derive a system of coupled differential equations that describe the nonlinear waves in conformal Israel-Stewart theory. In Section V we solve these equations numerically and we finish with our conclusions and outlook. Throughout this study we use natural units $\hbar=c=k_{B}=1$ and a mostly minus metric $g_{\mu\nu}=\textrm{diag}(+,-,-,-)$.

\section{Second-order conformal hydrodynamic equations - Israel-Stewart Theory}

The energy-momentum tensor of a relativistic fluid is
\begin{equation}
T^{\mu\nu} = \varepsilon u^\mu u^\nu -p \Delta^{\mu\nu} +\pi^{\mu\nu}
\end{equation}
where $\varepsilon$ is the energy density, $p$ is the pressure, $u^\mu$ is the fluid 4-velocity $u^{\mu}=(\gamma,\gamma\vec{v})$ and $\gamma$ is the Lorentz factor $\gamma=(1-v^{2})^{-1/2}$ (hence, $u^{\mu}u_{\mu}=1$). The connection between $\varepsilon$ and $p$ defines the equation of state, which will be taken to be that of a conformal fluid, $\varepsilon=3p$. The entropy density is then $s=\kappa T^3$, where $\kappa$ is a numerical coefficient. Furthermore, the projection operator orthogonal to the fluid velocity is given by $\Delta^{\mu\nu} \equiv g^{\mu\nu}-u^{\mu}u^{\nu}$. The shear stress tensor $\pi^{\mu\nu}$ is a symmetric (and traceless) tensor that is orthogonal to the flow $u_\mu \pi^{\mu\nu}=0$ (i.e, the Landau frame \cite{land}). Besides the energy-momentum conservation equations $\partial_\mu T^{\mu\nu}=0$, or in explicit form,
\begin{equation}
D\varepsilon + (\varepsilon+p)\theta-\pi^{\mu\nu}\,\sigma_{\mu\nu}=0
\label{f1}
\end{equation}
\begin{equation}
(\varepsilon+p)Du^{\alpha} - \nabla_{\perp}^{\alpha} \, p+\Delta^{\alpha}_{\nu}\partial_{\mu}
\pi^{\mu\nu}=0\,
\label{f2}
\end{equation}
the simplified set of the conformal IS equations \cite{brsss} that defines the dynamics of the shear stress tensor $\pi^{\mu\nu}$ used here are \cite{hugo}
\begin{equation}
{\tau}_{\pi} \Big(\Delta^{\mu}_{\alpha}\Delta^{\nu}_{\beta}D\pi^{\alpha \beta}+{\frac{4}{3}}
\pi^{\mu\nu} \theta \Big)+\pi^{\mu\nu}=2\eta\sigma^{\mu\nu}
\label{f3}
\end{equation}
where the operators are written in shorthand notation by $D \equiv u^{\mu}\partial_{\mu}$, which is the comoving derivative, $\nabla_{\perp}^{\alpha} \equiv \Delta^{\alpha\mu} \partial_{\mu}$ is the derivative orthogonal to the 4-velocity and the expansion rate is $\theta \equiv \partial^{\mu}u_{\mu}$. The shear tensor is defined as $\sigma^{\mu\nu}\equiv \Delta^{\mu\nu\alpha\beta}\partial_{\alpha}u_{\beta}$
using the doubly symmetric and traceless projection operator
$\Delta^{\mu\nu\alpha\beta} \equiv (\Delta^{\mu\alpha}\Delta^{\nu\beta}+\Delta^{\mu\beta}\Delta^{\nu\alpha})/2-
\Delta^{\mu\nu}\Delta^{\alpha\beta}/3$. Also, note that in a conformal fluid $\tau_\pi \sim 1/T$ and $\eta \sim T^3$. Therefore, in Israel-Stewart's theory the dynamical variables are the usual hydrodynamical quantities $\varepsilon$ and $u^\mu$ together with the shear stress tensor $\pi^{\mu\nu}$ (see the Appendix for a discussion about the components of the shear stress tensor that contribute in our analysis).  

Effects from bulk viscosity (see, for instance, \cite{Noronha-Hostler:2013gga,Noronha-Hostler:2014dqa}) or from additional conserved charges are not taken into account here. Furthermore, even though the equation of motion for $\pi^{\mu\nu}$ derived from kinetic theory contains many more terms than the ones used here \cite{denicolPRD}, in this paper we shall focus on the simplest set of equations that can still describe a causal (and stable) conformal dissipative fluid.

\section{Linearized wave equations}

In order to  obtain the simplest wave equation for a small perturbation in the fluid around equilibrium, one can resort to the formalism known as the
``{\it linearization formalism}'' \cite{hidro,nrev,fgn13}, in which one performs the following expansions of the energy
density, pressure, shear stress tensor, and fluid 4-velocity around their respective equilibrium configuration values (for simplicity, here we take the sound wave disturbances in the ``x" direction)
$$
\varepsilon(x,t)=\varepsilon_{0}+\delta{\varepsilon(x,t)}\,,  \hspace{0.8cm}
\tau_\pi(x,t)=\tau_\pi^{0}+\delta \tau_\pi(x,t) \,,  \hspace{0.8cm} \eta(x,t)=\eta_0+\delta \eta(x,t)
$$
\begin{equation}
\pi^{xx}= \delta \pi^{xx}(x,t)
\hspace{0.9cm} \textrm{and}  \hspace{0.9cm}
u^{\mu}(x,t)=(1,0,0,0)+(0,\delta u^{x}(x,t),0,0)
\label{hatvaress}
\end{equation}
In Eq.\ (\ref{hatvaress}), `` $\delta$ '' denotes a small deviation from equilibrium.
After inserting the expansions (\ref{hatvaress}) in Eqs.\ (2)-(4), the linearization approximation is performed by neglecting the
$\mathcal{O}({\delta^{n}})$ terms for $n\geq 2$ in the resulting equations (see the Appendix for a discussion about the different components of the shear stress tensor).

Linearizing the three equations (\ref{f1}) to (\ref{f3}), using that
${c_{s}}^{2}=dp/d\varepsilon=1/3$, we find
\begin{equation}
{\frac{\partial }{\partial t}}\delta \varepsilon+
{\frac{4}{3}}\varepsilon_{0}\,{\frac{\partial }{\partial x}}\delta u^{x}=0
\label{f1l}
\end{equation}
\begin{equation}
{\frac{4}{3}}\varepsilon_{0}\,{\frac{\partial }{\partial t}}\delta u^{x}+
{\frac{1}{3}}{\frac{\partial }{\partial x}}\delta \varepsilon+
{\frac{\partial }{\partial x}}\delta \pi^{xx}=0
\label{f2l}
\end{equation}
\begin{equation}
\tau_{\pi}^0{\frac{\partial}{\partial t}}\delta \pi^{xx}+\delta \pi^{xx}+
{\frac{4}{3}}\eta_0 {\frac{\partial }{\partial x}} \delta u^{x}=0\,,
\label{f3l}
\end{equation}
respectively.

Inserting (\ref{f2l}) and its time derivative into the spatial derivative of
(\ref{f3l}) we find
\begin{equation}
\tau_{\pi}^0 \Bigg[-{\frac{4}{3}}\varepsilon_{0}\,{\frac{\partial^{2}}{\partial t^{2}}}\delta u^{x}-{\frac{1}{3}}{\frac{\partial }{\partial t}}{\frac{\partial }{\partial x}}\delta \varepsilon \Bigg]-{\frac{4}{3}}\varepsilon_{0}\,{\frac{\partial }{\partial t}}\delta u^{x}-
{\frac{1}{3}}{\frac{\partial }{\partial x}}\delta \varepsilon
+{\frac{4}{3}}\eta_0 {\frac{\partial^{2}}{\partial x^{2}}} \delta u^{x}=0\,.
\label{prewe}
\end{equation}
Calculating the time and spatial derivative of (\ref{f1l}) and inserting these results in the  spatial derivative of (\ref{prewe}), together with the Gibbs relation for the background $4\varepsilon_{0}/3=T_{0}s_{0}$, we obtain the following wave equation
\begin{equation}
{\frac{\partial^{2}}{\partial x^{2}}}\delta\varepsilon
-3{\frac{\partial^{2}}{\partial t^{2}}}\delta\varepsilon
-3{\tau_{\pi}^0}{\frac{\partial^{3}}{\partial t^{3}}}\delta\varepsilon=
- \bigg({\frac{3 \chi}{T_{0}\, }} +{\tau_{\pi}^0} \bigg)  {\frac{\partial}{\partial t}}{\frac{\partial^{2} }{\partial x^{2}}}\delta\varepsilon
\label{lossy}
\end{equation}
with dissipation and relaxation time effects. Equation (\ref{lossy}) is sometimes known as the {\it lossy wave equation} \cite{holm1,holm2}.
The dimensionless coefficient $\chi$ is given by
\begin{equation}
\chi={\frac{4}{3}}{\frac{\eta_0}{s_{0}}}\,.
\label{viscoeflossy}
\end{equation}
In the limit $\tau_{\pi}^0 \to 0$ one recovers the linear wave equation for the viscous fluid described by the relativistic Navier-Stokes theory \cite{nrev,fgn13}. Also, setting $\tau_{\pi}^0=\chi=0$, one obtains the linear wave equation for the ideal relativistic fluid \cite{nrev,fgn13}.

\subsubsection{Stability and causality}

To study some properties of (\ref{lossy}) we consider a plane wave Ansatz for all the disturbances, e.g.,
\begin{equation}
\delta \varepsilon(x,t)=\mathcal{A} \, e^{i(kx-\omega t)}\,,
\label{ansatz}
\end{equation}
which yields the following dispersion relation \cite{shipu2}
\begin{equation}
\omega^{2}={\frac{k^{2}}{3}}{\frac{(1-i\Lambda\omega)}{(1-i \tau_{\pi}^0 \omega)}}
\label{drlossy}
\end{equation}
where $\Lambda$ is given by
\begin{equation}
\Lambda \equiv \Bigg({\frac{3\chi}{T_{0}}}+\tau_{\pi}^0\Bigg)\,.
\label{drlossycons}
\end{equation}
For $\chi=0$ and $\tau_{\pi}^0=0$ the dispersion relation (\ref{drlossy}) becomes the
ideal fluid dispersion relation for a sound wave: $\omega^{2}=k^{2}/3$.  Setting only ${\tau_{\pi}}^0 \to 0$, (\ref{drlossy}) gives the
Navier-Stokes sound wave dispersion relation \cite{fgn13} $\omega^{2}=k^{2}/3-i\omega k^{2}\chi/T_{0}$.

Introducing the dimensionless variables
$\hat{\omega}=\omega/T_{0}$, $\hat{k}=k/T_{0}$ and $\hat{\tau}_{\pi}=T_{0}\,\tau_{\pi}^0$ , the dispersion relation (\ref{drlossy}) is rewritten as a dimensionless equation
to be solved for ${\hat{\omega}}$
\begin{equation}
-i \, \hat{\tau}_{\pi} \, {\hat{\omega}}^{3} +{\hat{\omega}}^{2}
+\Bigg(i\, \hat{k}^{2} \,\chi+{\frac{i}{3}}\,\hat{k}^{2}
\, \hat{\tau}_{\pi}   \Bigg) {\hat{\omega}}
-{\frac{\hat{k}^{2}}{3}}=0\,.
\label{drlossyad}
\end{equation}

We expect the hydrodynamic description to be meaningful for small values of $\hat{\omega}$ and $\hat{k}$. For completeness we extrapolate
our results to large values of $\hat{k}$, as motivated by \cite{acausal}. The study of the short wavelength limit is surely limited to phenomenological applications of fluid dynamics; however, since the relativistic Navier-Stokes theory is known to have numerical instabilities, we find it useful to check that the Israel-Stewart construction is free of any acausality and instability in this regime under linear perturbations. Our results from this section are consistent with the discussion presented in \cite{shipu1,shipu2}.

To study the stability and causality properties of (\ref{lossy}), we decompose the roots of (\ref{drlossyad}) in two components as in \cite{fgn13,shipu2}:
$\hat{\omega}=Re[\hat{\omega}]+i \, Im[\hat{\omega}]$ , where $Re[\hat{\omega}] \in \mathbb{R}$ and $Im[\hat{\omega}] \in \mathbb{R}$.  This decomposition also turns the solution (\ref{ansatz}) into
\begin{equation}
\delta \varepsilon(\hat{x},\hat{t})=\mathcal{A} \, e^{Im[\hat{\omega}]\hat{t}}e^{iRe[\hat{\omega}] \big(\hat{k}\hat{x}/Re[\hat{\omega}]-\hat{t}\big)}
\label{ansatza}
\end{equation}
and again, $\hat{x}=T_{0}\,x$ and $\hat{t}=T_{0}\,t$ are dimensionless quantities.
In (\ref{ansatza}) it is possible to identify the attenuation coefficient $Im[\hat{\omega}]$,
which dictates the stability properties of the disturbance, i.e., stable perturbations have $Im[\hat{\omega}]<0$.  The phase velocity $\hat{v}_{p}$ and the group velocity $\hat{v}_{g}$ are given by the following expressions
\begin{equation}
{\hat{v}_{p}}(\hat{k})={\frac{Re[\hat{\omega}]}{\hat{k}}}
\hspace{2cm} \textrm{and} \hspace{2cm}
{\hat{v}_{g}}(\hat{k})={\frac{dRe[\hat{\omega}]}{d\hat{k}}}
\label{phasegroup}
\end{equation}
and causality violation occurs if ${\hat{v}_{g}}$ diverges \cite{moyses}.

\begin{figure}[ht!]
\begin{center}
\subfigure[ ]{\label{fig1a}
\includegraphics[width=0.485\textwidth]{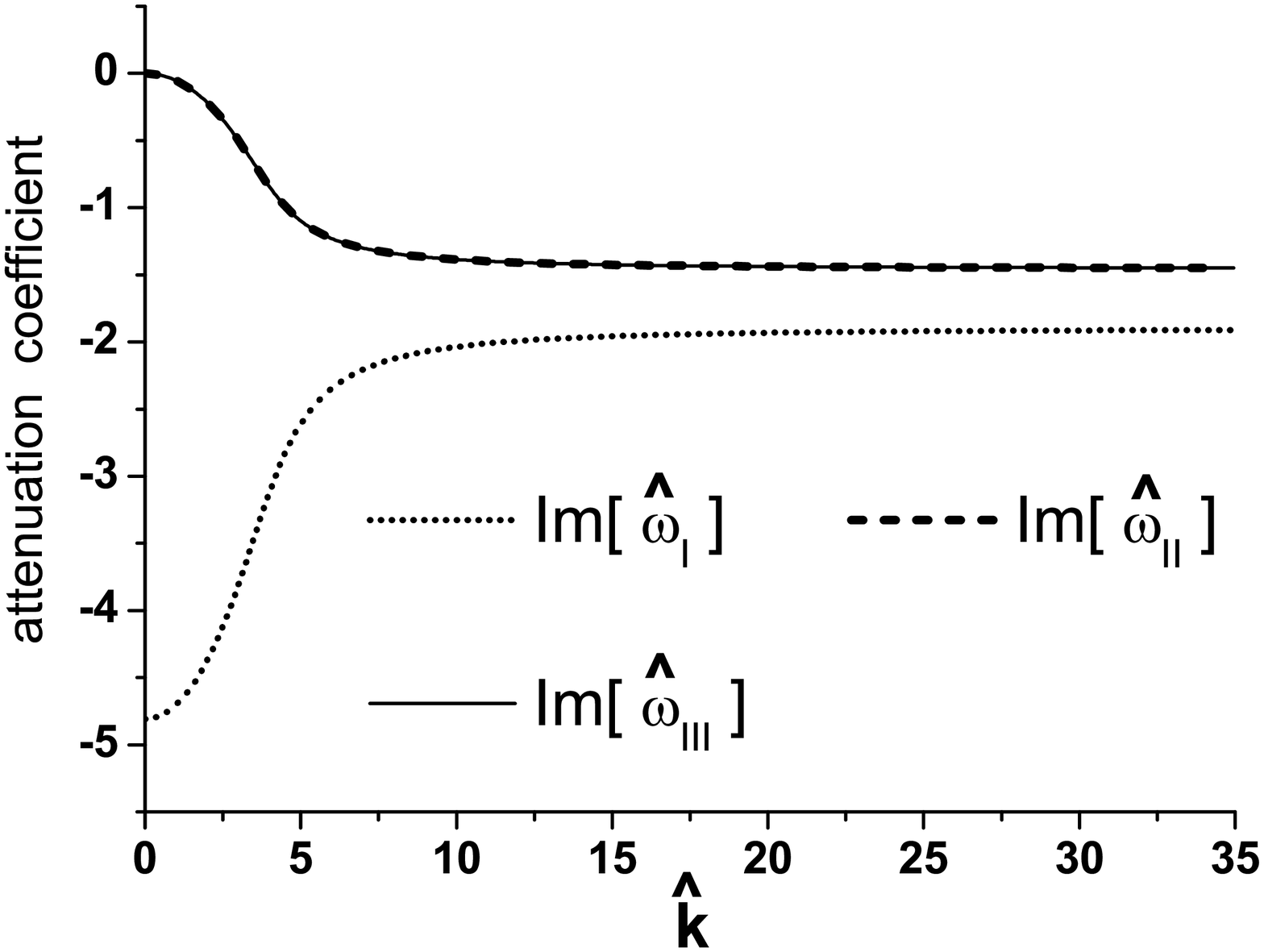}}
\subfigure[ ]{\label{fig1b}
\includegraphics[width=0.485\textwidth]{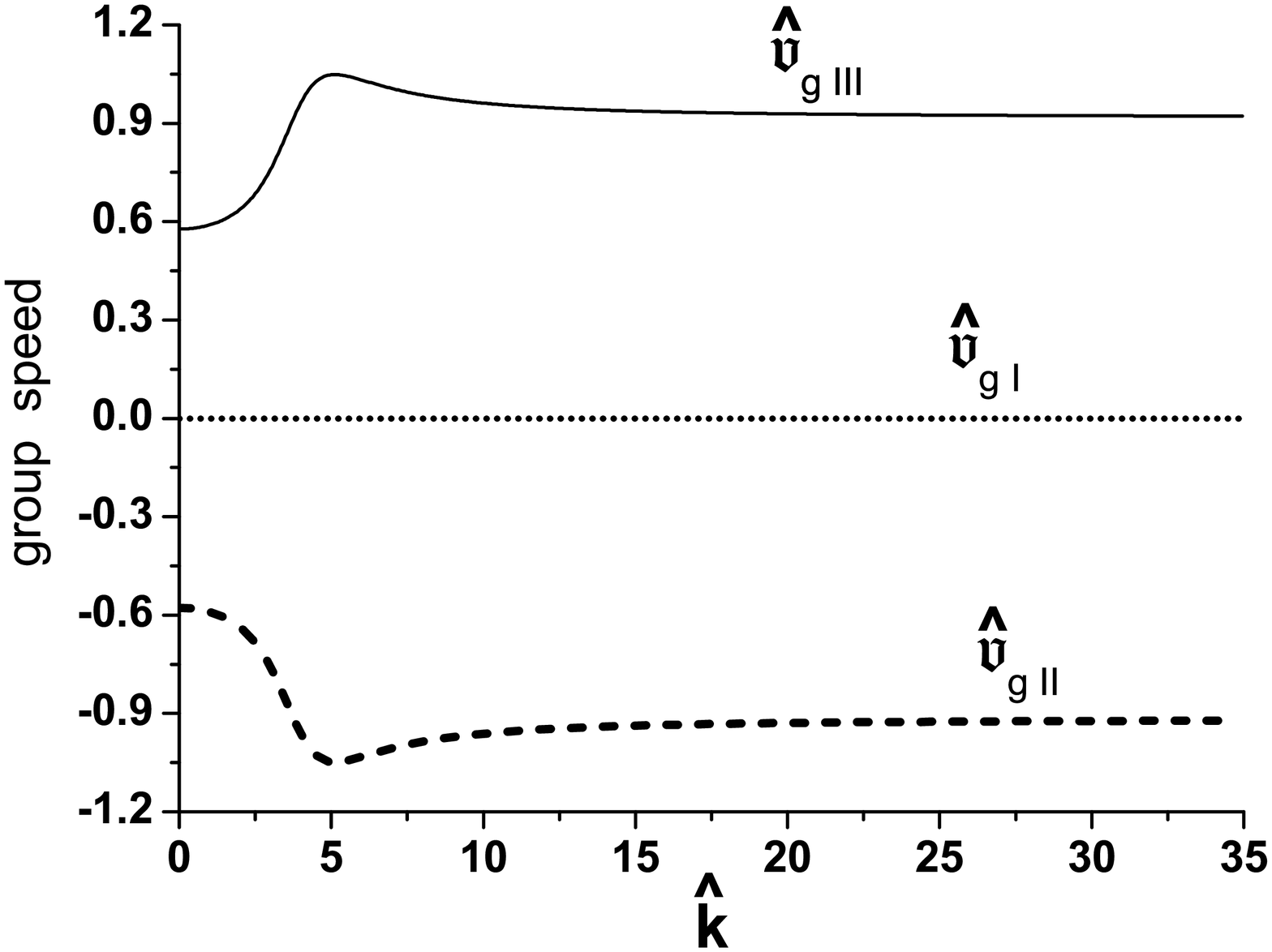}}
\end{center}
\caption{Stability and causality properties of disturbances around equilibrium for IS hydrodynamics (described by Eq.\ (\ref{lossy})) with transport coefficients from strongly-coupled $\mathcal{N}=4$ SYM.}
\label{fig1}
\end{figure}

In the analysis below we consider the coefficients of the strongly coupled $\mathcal{N}=4$ Supersymmetric Yang-Mills (SYM) fluid where $\eta_0/s_{0}=1/(4\pi)$ and $\hat{\tau}_{\pi}=[2-ln(2)]/(2\pi)$ \cite{brsss}. In Fig.\ \ref{fig1} we plot the group velocity ${\hat{v}_{g}}$ and the attenuation coefficient $Im[\hat{\omega}]$ for the three roots $\hat{\omega}_{I}$, $\hat{\omega}_{II}$ and $\hat{\omega}_{III}$ of (\ref{drlossyad}). In Fig.\ \ref{fig1a} one can clearly notice that the three modes are stable since the imaginary parts of the modes are always negative.  In Fig.\ \ref{fig1b}
there is no causality violation since there is no divergence as $\hat{k}$ increases and the group velocity is bounded by unity for large values of $\hat{k}$. The influence of the group velocity in the causal aspects of wave propagation comes from solving the full propagator in configuration space, which follows from the integral in momentum space.  The resulting propagator should be a function defined only inside the corresponding lightcone. If for large $\hat{k}$ the group velocity is at maximum unity, then this condition is satisfied \cite{shipu2}. This figure shows that the linear sound wave disturbances around thermodynamical equilibrium in 2nd order hydrodynamics (with the transport coefficients of strongly-coupled $\mathcal{N}=4$ SYM) are causal and stable. A similar study can be done for the shear channel \cite{shipu2}.

The three modes of the IS theory with strongly-coupled transport coefficients shown in Fig.\ \ref{fig1} can be expanded in powers of $\hat{k}$.
Since we are considering dimensionless variables, the limit of small $\hat{k}$ and $\hat{\omega}$ reveals the infrared behavior of the theory, i.e., the behavior of the modes with $\hat{\omega},\hat{k} << 1$.  Hydrodynamics can be considered as an effective theory at low energies and, thus, it should be well defined for small $\hat{k}$. We obtain the following relations
\begin{equation}
{\hat{\omega}}_{I}={\frac{\hat{k}}{\sqrt{3}}}-{\frac{i}{2}}{\chi }\hat{k}^{2}
-{\frac{\sqrt{3}}{6}}{\chi}\Bigg({\frac{3}{4} \chi}- \hat{\tau}_{\pi}\Bigg)\hat{k}^{3}
-{\frac{i}{6}}{\chi}\hat{\tau}_{\pi}\Bigg(3 \chi- \hat{\tau}_{\pi} \Bigg)\hat{k}^{4}
+\mathcal{O}(\hat{k}^{5})
\label{firstroot}
\end{equation}
\begin{equation}
{\hat{\omega}}_{II}=-{\frac{\hat{k}}{\sqrt{3}}}-{\frac{i}{2}}{\chi }\hat{k}^{2}
+{\frac{\sqrt{3}}{6}}{\chi}\Bigg({\frac{3}{4} \chi}- \hat{\tau}_{\pi}\Bigg)\hat{k}^{3}
-{\frac{i}{6}}{\chi}\hat{\tau}_{\pi}\Bigg(3 \chi- \hat{\tau}_{\pi} \Bigg)\hat{k}^{4}
+\mathcal{O}(\hat{k}^{5})
\label{secondroot}
\end{equation}
\begin{equation}
{\hat{\omega}}_{III}=-{\frac{i}{\hat{\tau}_{\pi}}}+i\,{}{\chi}\hat{k}^{2}
+i{\frac{\chi}{3}}\hat{\tau}_{\pi}\Bigg(3 \chi- \hat{\tau}_{\pi} \Bigg)\hat{k}^{4}
+\mathcal{O}(\hat{k}^{5})
\label{thirdroot}
\end{equation}
The first two modes describe sound waves at small momenta while the 3rd mode is a non-hydrodynamic mode. For small $\hat{k}$, the relaxation coefficient $\hat{\tau}_{\pi}$ first appears in the sound wave modes multiplied by $ 3 \chi/4$ ($\eta_0/s_{0}$) in the ${\hat{k}}^{3}$ term \cite{brsss}. However, note that $\hat{\tau}_{\pi}$ already appears at zeroth order in $\hat{k}$ in $\hat{\omega}_{III}$ and it clearly defines a microscopic relaxation time scale \cite{noronhaPRD}.

It is important to notice that in this regime the non-hydrodynamic mode $\hat{\omega}_{III}$ should only play a role when $\hat{k} \approx 1$, where its value becomes comparable to the other modes.  The physical interpretation is that in the low energy limit one should only care about the hydrodynamic modes.  However, notice that this mode is stable for any $\hat{k}$, which implies that Israel-Stewart theory is well-defined mathematically in the short wavelength limit.

We followed closely the study performed in \cite{shipu2} for the Israel-Stewart theory where the dispersion relation (\ref{drlossy}) and the decomposition of its modes in real and imaginary parts were considered in the large and small wavenumber limits. In \cite{shipu2} the authors suggested that
the problem of acausality and instability are correlated in relativistic dissipative hydrodynamics.

For the modes (\ref{firstroot}), (\ref{secondroot}) and (\ref{thirdroot}), the group velocity and
the attenuation coefficient at small $\hat{k}$ are
\begin{equation}
{\hat{v}_{g \, I}}=-{\hat{v}_{g \, II}} = {\frac{1}{\sqrt{3}}}-{\frac{\sqrt{3}}{2}}{\chi}\Bigg({\frac{3}{4} \chi}- \hat{\tau}_{\pi}\Bigg)\hat{k}^{2} + \mathcal{O}({\hat{k}^{4}})
\hspace{1.5cm} \textrm{and} \hspace{1.5cm}
{\hat{v}_{g \, III}}=0
\end{equation}
$$
Im[\hat{\omega}_{I}]=Im[\hat{\omega}_{II}] = -\,{\frac{\chi}{2}} \hat{k}^{2}
-{\frac{\chi}{6}} \hat{\tau}_{\pi}\Bigg(3 \chi- \hat{\tau}_{\pi} \Bigg)\hat{k}^{4}
+ \mathcal{O}({\hat{k}^{5}}) <0
\hspace{1.5cm} \textrm{and} \hspace{1.5cm}
$$
\begin{equation}
Im[\hat{\omega}_{III}] = -\Bigg({\frac{1}{\hat{\tau}_{\pi}}}-\, \chi \hat{k}^{2} \Bigg)+{\frac{\chi}{3}}\hat{\tau}_{\pi}\Bigg(3{\chi}- \hat{\tau}_{\pi} \Bigg)\hat{k}^{4}
+ \mathcal{O}({\hat{k}^{5}})<0\,.
\label{atts}
\end{equation}

\section{Nonlinear wave equations in Conformal Israel-Stewart theory}

The effects from a relaxation timescale $\hat{\tau}_{\pi}$ have not yet been studied in the context of nonlinear wave propagation.  In order to investigate its effects in the study of nonlinear waves, we shall use the Reductive Perturbation Method (RPM) \cite{davidson,rpm,leblond,loke}.  The RPM was used to study nonlinear waves in
relativistic and non-relativistic hydrodynamics in \cite{nrev,fgn13,we10,nos2012}.  Our goal in this section is to find the nonlinear
wave equation that governs the perturbation of the energy density in a hot dissipative and causal fluid described by IS hydrodynamics.

\subsection{Reductive Perturbation Method}

With the RPM  we can derive the nonlinear wave equation for perturbations in a fluid performing the following set of operations \cite{nrev,fgn13}:

\vspace{0.5cm}

$(a)$  Rewrite Eqs.\ (\ref{f1}), (\ref{f2}), and (\ref{f3}) using the following dimensionless variables
\begin{equation}
\hat \varepsilon(x,t)={\frac{\varepsilon(x,t)}{\varepsilon_{0}}} \hspace{0.2cm} \textrm{,}   \hspace{0.25cm}
{\hat v}_{x}(x,t)={\frac{v_{x}(x,t)}{c_{s}}}=\sqrt{3} \,\, v_{x}(x,t)
\hspace{0.8cm} \textrm{and}   \hspace{0.8cm}
\hat \pi^{xx}(x,t)={\frac{\pi^{xx}(x,t)}{p_0}}
\label{hatvarers}
\end{equation}

\vspace{0.5cm}

$(b)$ Change the coordinates in Eqs.\ (\ref{f1}), (\ref{f2}), and (\ref{f3}) from $(x,t)$ to the $(X,Y)$ space defined by the ``stretched coordinates'' \cite{davidson,rpm,leblond,loke}
\begin{equation}
X=\sigma^{1/2}{\frac{(x-{c_{s}}t)}{L}}=\sigma^{1/2}{\frac{1}{L}}\bigg(x-{\frac{t}{\sqrt{3}}}\bigg)
\hspace{1.5cm} \textrm{and}   \hspace{1.5cm}
Y=\sigma^{3/2}{\frac{{c_{s}}t}{L}}=\sigma^{3/2}{\frac{t}{\sqrt{3}L}}
\label{xtst}
\end{equation}
where $L$ is a characteristic length scale of the problem, which will be simplified in the final expressions, and $\sigma$ is a small ($0 < \sigma < 1$), dimensionless expansion parameter. We also change the shear viscosity coefficient and the relaxation time to the $(X,Y)$ space in the following way \cite{stvis1,stvis2}
\begin{equation}
\eta=\sigma^{1/2} \, \tilde{\eta}
\hspace{2cm} \textrm{and}   \hspace{2cm}
\tau_{\pi}=\sigma^{1/2} \, \tilde{\tau}_{\pi}\,.
\label{stv}
\end{equation}
We refer the reader to the Appendix for the details. While the scaling of $\eta$ with $\sigma$ was known in literature \cite{stvis1,stvis2}, the proposed scaling of $\tau_\pi$ with $\sigma$ used here is new and it is the simplest choice that is consistent with the sound mode dispersion relation (\ref{firstroot}) and (\ref{secondroot}).

\vspace{0.5cm}

$(c)$ Expand the variables in Eq.\ (\ref{hatvarers}) around their equilibrium values
\begin{equation}
\hat\varepsilon={\frac{\varepsilon}{\varepsilon_{0}}}=1+\sigma \varepsilon_{1}+ \sigma^{2} \varepsilon_{2}+
\sigma^{3} \varepsilon_{3}+\dots
\label{enerdenexp}
\end{equation}
\begin{equation}
{\hat v}_{x}={\frac{v_{x}}{c_{s}}}=\sigma {v_{1}}+ \sigma^{2} {v_{2}}+ \sigma^{3} {v_{3}}+\dots
\label{vexp}
\end{equation}
and
\begin{equation}
\hat \pi^{xx}={\frac{\pi^{xx}}{p_0}}=\sigma {\pi^{xx}_{1}}+ \sigma^{2} {\pi^{xx}_{2}}+ \sigma^{3} {\pi^{xx}_{3}}+\dots\,.
\label{piexp}
\end{equation}

After the expansions, we organize the resulting equations in powers of $\sigma$, neglecting terms with powers greater than $\sigma^{3}$.  In the usual RPM method, only terms proportional to $\sigma$ and $\sigma^{2}$ are kept.  However, the linear hydrodynamical modes (\ref{firstroot}) and (\ref{secondroot}) show that relaxation effects may appear only in the next order of the usual expansion, i.e., at order $\hat{k}^{3}$.  For this reason we consider the $\sigma$ expansion up to $\mathcal{O}({\sigma^{3}})$ terms to study relaxation effects in nonlinear waves.

\vspace{0.5cm}

$(d)$ By solving the system of algebraic equations: $\sigma^{1/2}\{\dots\}=0$, $\dots$, $\sigma^{3}\{\dots\}=0$ obtained in the step $(c)$, it is possible to
find the system of wave equations in the $(X,Y)$ space.  Such system may be transformed back to the $(x,t)$ coordinates through the stretching transformations
(\ref{xtst}) and (\ref{stv}) yielding the final system of nonlinear wave equations for the perturbations in the energy density.

\subsection{Nonlinear wave equations}

The set of differential equations obtained from the RPM method is given by
\begin{equation}
{\frac{\partial}{\partial \hat{t}}}\hat{\varepsilon}_{1}+{\frac{1}{\sqrt{3}}}\,{\frac{\partial}{\partial \hat{x}}}
\hat{\varepsilon}_{1}
+{\frac{1}{2\sqrt{3}}}\,\hat{\varepsilon}_{1}{\frac{\partial}{\partial \hat{x}}}\hat{\varepsilon}_{1}
={\frac{\chi}{2}} \, {\frac{\partial^{2} }{\partial \hat{x}^{2}}}\hat{\varepsilon}_{1}
\label{burgXTcad}
\end{equation}
and
$$
{\frac{\partial}{\partial \hat{t}}}\hat{\varepsilon_{2}}+
{\frac{1}{\sqrt{3}}}\,{\frac{\partial}{\partial \hat{x}}}
\hat{\varepsilon}_{2}
+{\frac{1}{2\sqrt{3}}}\,\hat{\varepsilon}_{1}{\frac{\partial}{\partial \hat{x}}}\hat{\varepsilon}_{2}
-{\frac{\chi}{2}} \,{\frac{\partial^{2} }{\partial {\hat{x}}^{2}}}\hat{\varepsilon}_{2}
+{\frac{1}{2\sqrt{3}}}\,\hat{\varepsilon}_{2}{\frac{\partial}{\partial \hat{x}}}\hat{\varepsilon}_{1}
+{\frac{\chi}{4}}\,
\hat{\varepsilon}_{1}{\frac{\partial^{2}}{\partial {\hat{x}}^{2}}}\hat{\varepsilon}_{1}
$$
\begin{equation}
+{\frac{1}{4}}\,
\hat{\varepsilon}_{1}{\frac{\partial}{\partial \hat{t}}}\hat{\varepsilon}_{1}
+{\frac{1}{4\sqrt{3}}}\,
\hat{\varepsilon}_{1}{\frac{\partial}{\partial \hat{x}}}\hat{\varepsilon}_{1}
+{\frac{\chi}{2}}
\Bigg[{\frac{\chi \, \sqrt{3}}{4}}
 -{\frac{\hat{\tau}_{\pi}}{\sqrt{3}}}\Bigg]{\frac{\partial^{3} }{\partial {\hat{x}}^{3}}}\hat{\varepsilon}_{1}=0\,,
\label{weeps2onlyad}
\end{equation}
where $\hat{\varepsilon}_1\equiv \sigma \varepsilon_1$ and $\hat{\varepsilon}_2\equiv \sigma^2 \varepsilon_2$.  The details of the calculations and assumptions needed to derive these equations are presented in the Appendix. Also, given the solution of (29) and (30), one is also able to study the behavior of $\hat\pi^{xx}$. However, in this paper we shall focus on the energy density disturbance and leave a detailed study of the shear stress tensor in this approach for future work.

We emphasize that the Burgers' equation (\ref{burgXTcad}) for the first order energy perturbation $\hat{\varepsilon}_{1}$ does not contain relaxation effects and, thus, it is the same both in Navier-Stokes and in Israel-Stewart theory.  This feature has lead us to consider perturbations up to third order in energy density and fluid velocity.  This provides the first equation where the relaxation time coefficient appears: Eq.\ (\ref{weeps2onlyad}) for $\hat{\varepsilon}_{2}$.

\section{Numerical results and discussion}

An analytical solution of the Burgers' equation (\ref{burgXTcad}) can be obtained by the
{\it{hyperbolic tangent expansion method}} \cite{fgn13} and its variants.  However, it is not possible to find a finite solution after substituting the analytical solution for $\hat{\varepsilon}_{1}$ into (\ref{weeps2onlyad}) to solve it for $\hat{\varepsilon}_{2}$. We have thus proceeded to solve (\ref{burgXTcad}) and (\ref{weeps2onlyad}) numerically.

\subsection{Soliton initial profile}

\begin{figure}[ht!]
\begin{center}
\subfigure[ ]{\label{fig2a}
\includegraphics[width=0.485\textwidth]{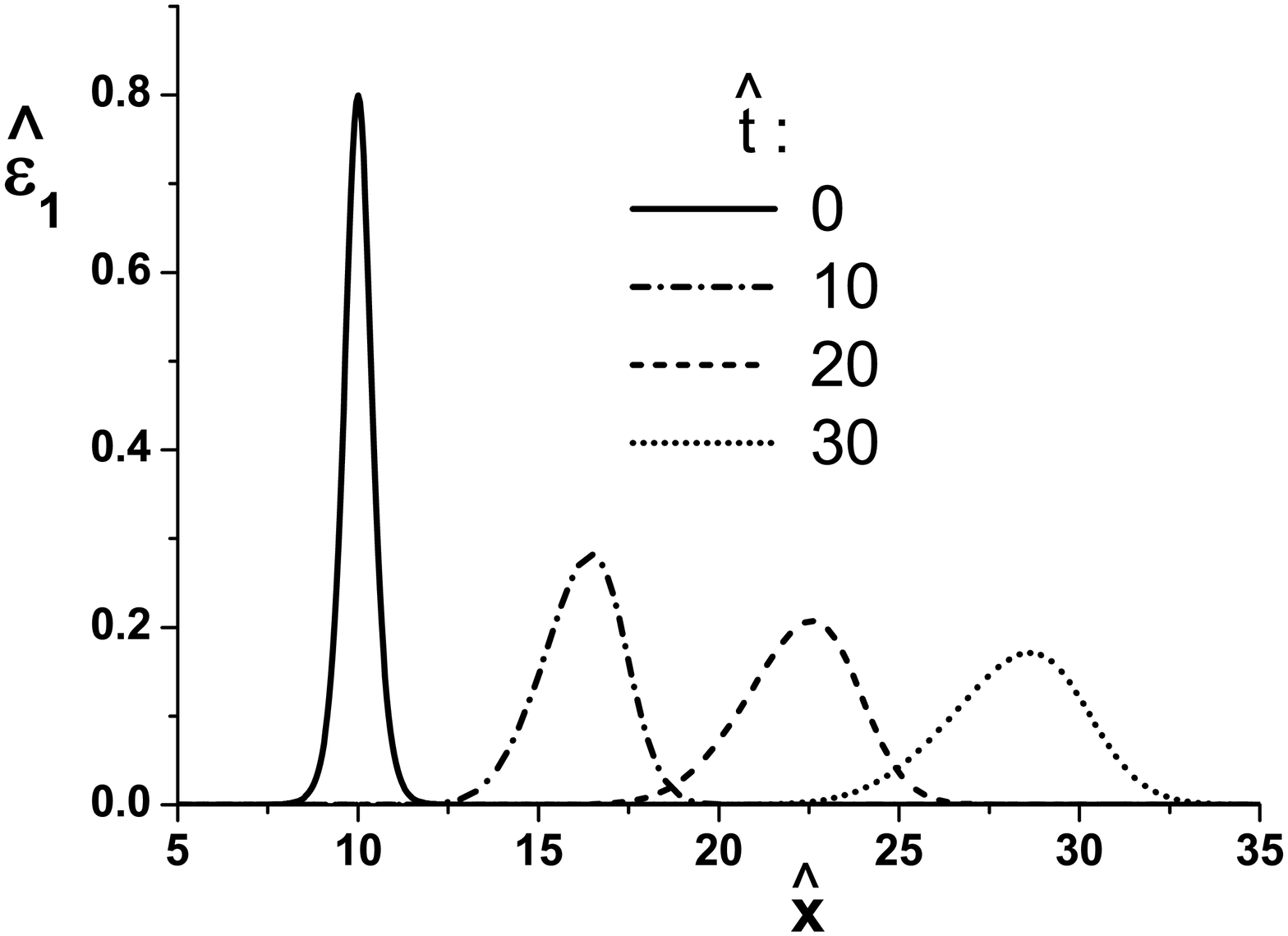}}
\subfigure[ ]{\label{fig2b}
\includegraphics[width=0.485\textwidth]{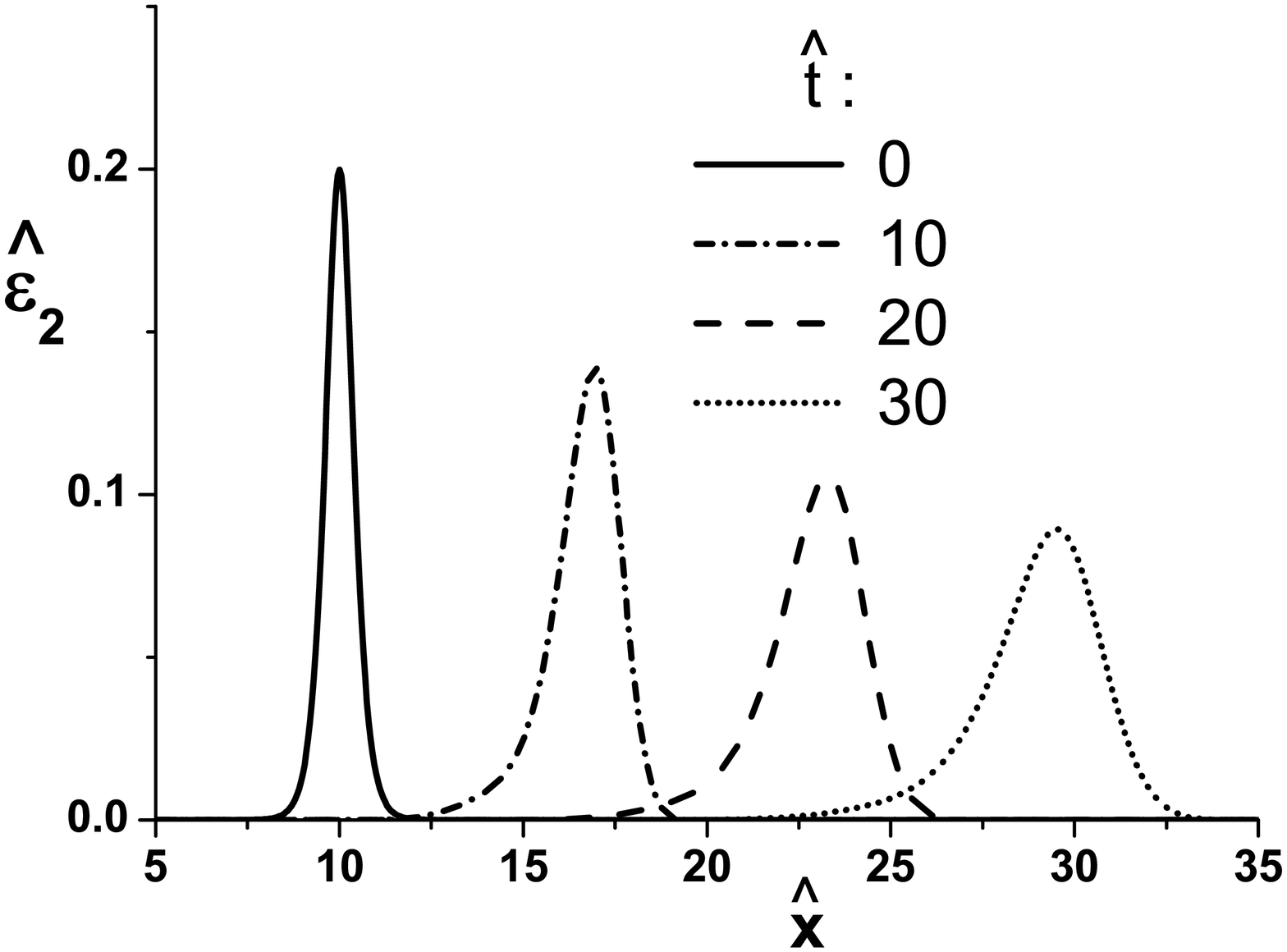}}\\
\subfigure[ ]{\label{fig2c}
\includegraphics[width=0.485\textwidth]{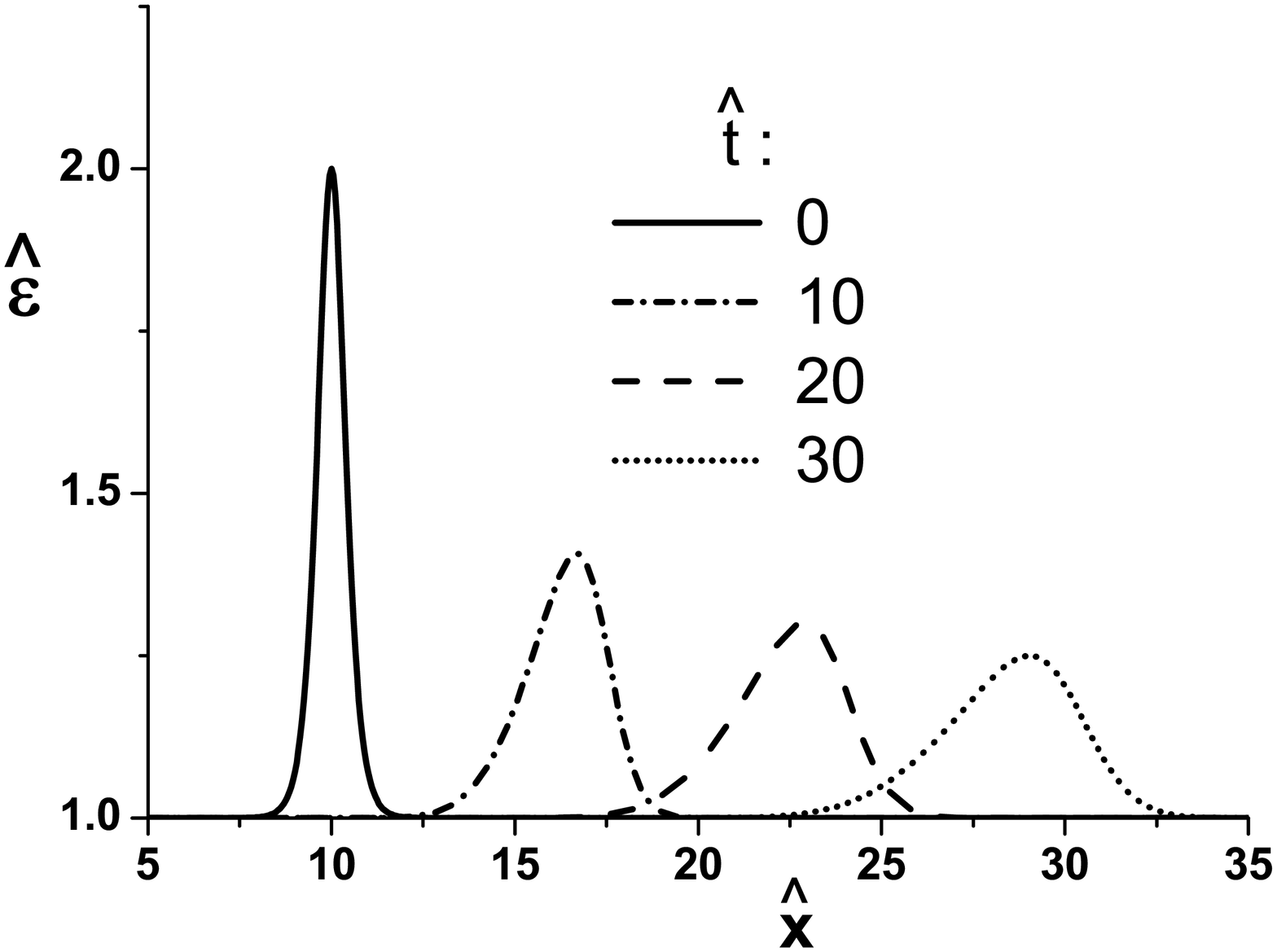}}
\end{center}
\caption{Numerical solutions for the energy density disturbances in the nonlinear regime in Eqs.\ (29) (fig. \ref{fig2a}) and (30) (fig. \ref{fig2b}) for $\eta_0/s_{0}=1/(4\pi)$ and $\hat{\tau}_{\pi}=[2-ln(2)]/(2\pi)$. The initial conditions are (\ref{sle1}) and (\ref{sle2})
with $A_{1}=0.8$, $A_{2}=0.2$ and $B_{1}=B_{2}=0.5$ . The fig. \ref{fig2c} shows the complete energy density perturbation
$\hat\varepsilon=1+\hat{\varepsilon}_{1}+ \hat{\varepsilon}_{2}$ .  The perturbations survive despite the dissipative effects.}
\label{fig2}
\end{figure}

Several different sets of parameters and initial profiles are considered in this study. Starting with the following typical strong coupling parameters $3\chi/4=\eta_0/s_{0}=1/(4\pi)$ and $\hat{\tau}_{\pi}=[2-ln(2)]/(2\pi)$ \cite{brsss}, we show the propagation of nonlinear waves in Fig.\ \ref{fig2}. We start by solving (\ref{burgXTcad}) with the following initial condition
\begin{equation}
\hat{\varepsilon}_{1}(\hat{x},0)= A_{1} \ sech^{2}\bigg(\frac{\hat{x}}{B_{1}}\bigg)
\label{sle1}
\end{equation}
and inserting the obtained numerical solution of (\ref{burgXTcad}) into (\ref{weeps2onlyad}) with the initial profile for
$\hat{\varepsilon}_{2}$
\begin{equation}
\hat{\varepsilon}_{2}(\hat{x},0)= A_{2} \ sech^{2}\bigg(\frac{\hat{x}}{B_{2}}\bigg)\,.
\label{sle2}
\end{equation}

The first case in Fig.\ \ref{fig2} corresponds to $A_{1}=0.8$, $A_{2}=0.2$ and $B_{1}=B_{2}=0.5$ . The numerical solution of (\ref{burgXTcad}), (\ref{weeps2onlyad}), and the total energy perturbation given by (\ref{enerdenexp}), $\hat\varepsilon=1+\hat{\varepsilon}_{1}+ \hat{\varepsilon}_{2}$, are shown in Fig.\ \ref{fig2}.  We notice that, in spite of the dissipative and relaxation effects, the perturbations still survive as time increases.

\begin{figure}[ht!]
\begin{center}
\subfigure[ ]{\label{fig3a}
\includegraphics[width=0.485\textwidth]{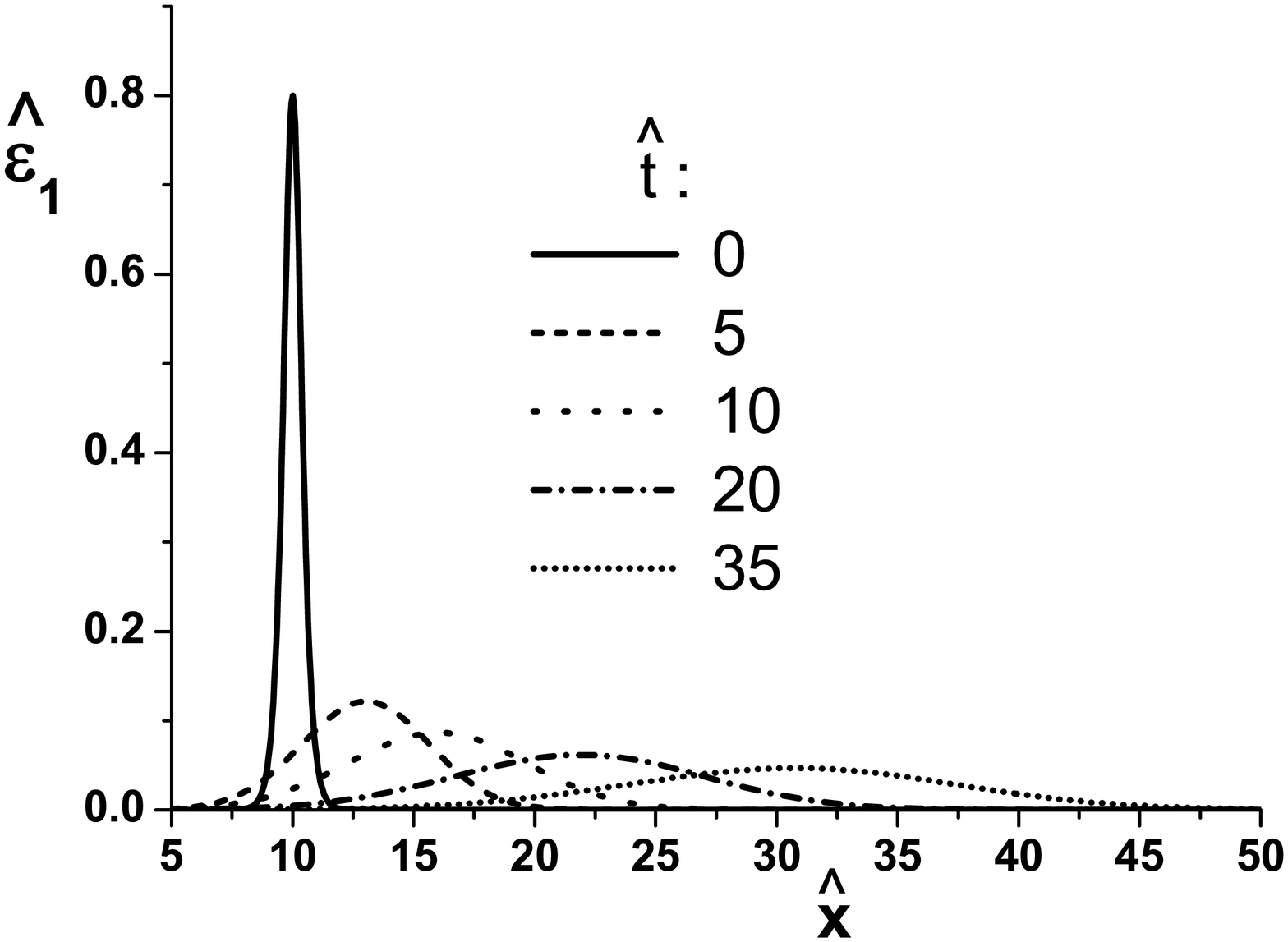}}
\subfigure[ ]{\label{fig3b}
\includegraphics[width=0.485\textwidth]{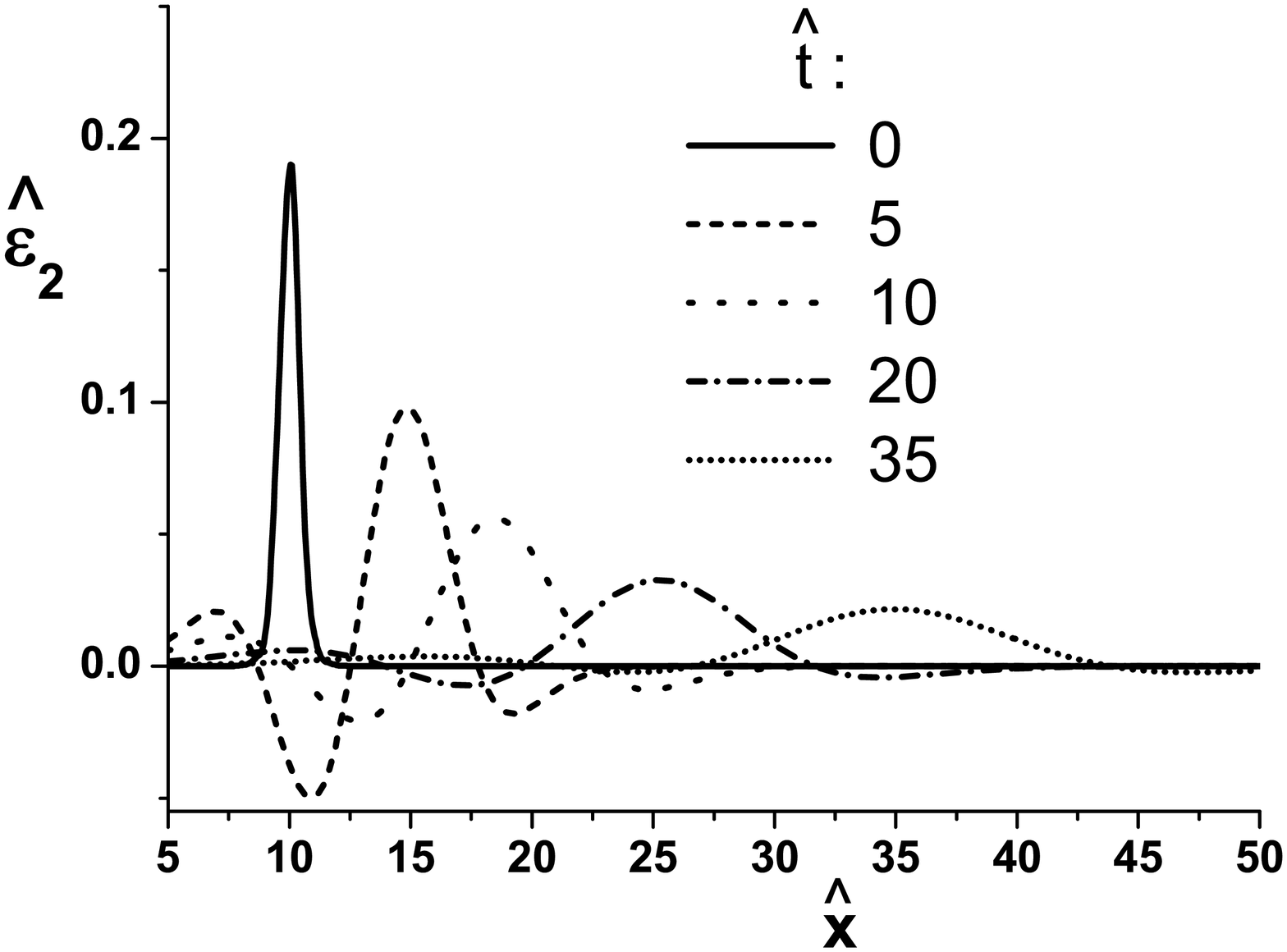}}\\
\subfigure[ ]{\label{fig3c}
\includegraphics[width=0.485\textwidth]{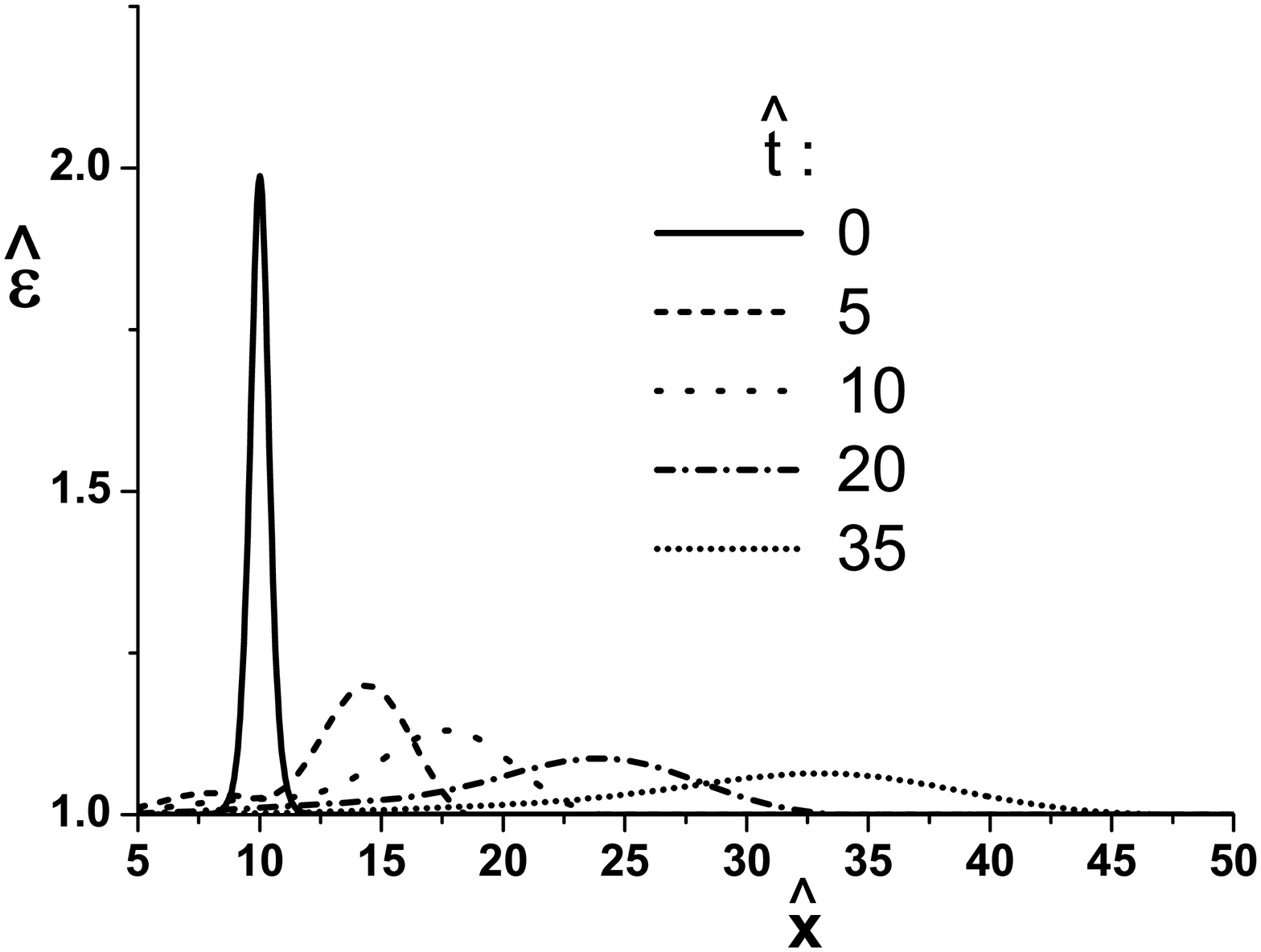}}
\end{center}
\caption{Numerical solutions for the energy density disturbances in the nonlinear regime in Eqs.\ (29) (fig. \ref{fig3a}) and (30) (fig. \ref{fig3b}) for $\eta_0/s_{0}=1$ and $\hat{\tau}_{\pi}=5\eta_0/s_0$. The initial conditions are (\ref{sle1}) and (\ref{sle2})
with $A_{1}=0.8$, $A_{2}=0.2$ and $B_{1}=B_{2}=0.5$ . The fig. \ref{fig3c} shows the complete energy density perturbation.  The perturbations do not survive due large dissipative effects.}
\label{fig3}
\end{figure}

In Fig.\ \ref{fig3} we show similar calculations as in Fig.\ \ref{fig2} but now considering large viscosity and relaxation time coefficients given respectively by  $\eta_0/s_{0}=1$ and $\hat{\tau}_{\pi}=5 \eta_0/s_0$, which is in the ballpark of kinetic theory calculations \cite{denicolPRD,noronhaPRD}. In Fig.\ \ref{fig3a} we obtain the expected result for the Burgers' equation with large viscosity: a strong dissipation of the initial pulse. In Fig.\ \ref{fig3b} we also obtain the same dissipation effect but at some intermediate time scales $\hat{t}=5$ to $\hat{t}=20$ there is also rarefaction.  The total perturbation does not survive for longer times and the perturbed fluid tends to recover the background configuration $\hat{\varepsilon}=1$ as time increases.

\begin{figure}[ht!]
\begin{center}
\subfigure[ ]{\label{fig4a}
\includegraphics[width=0.485\textwidth]{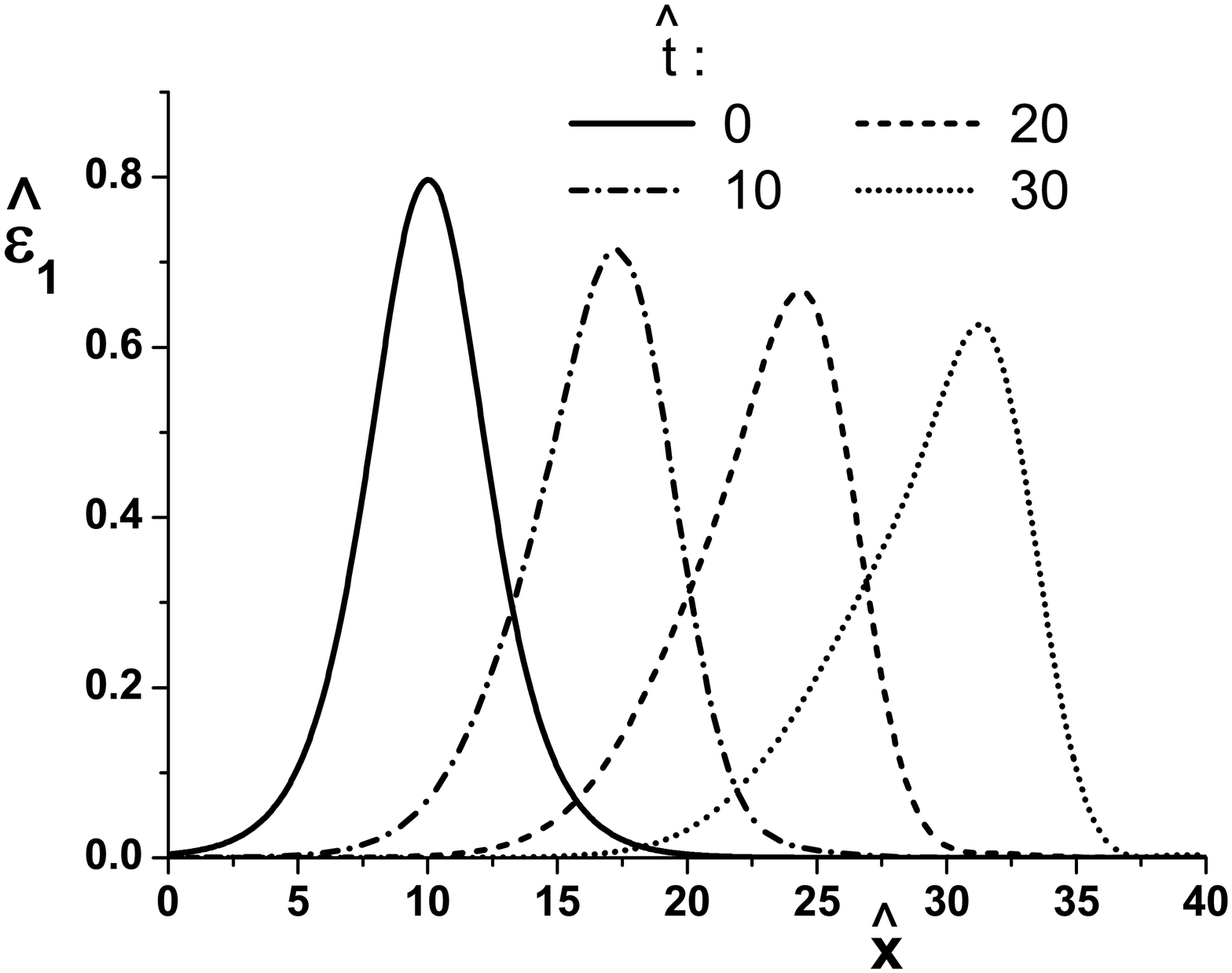}}
\subfigure[ ]{\label{fig4b}
\includegraphics[width=0.485\textwidth]{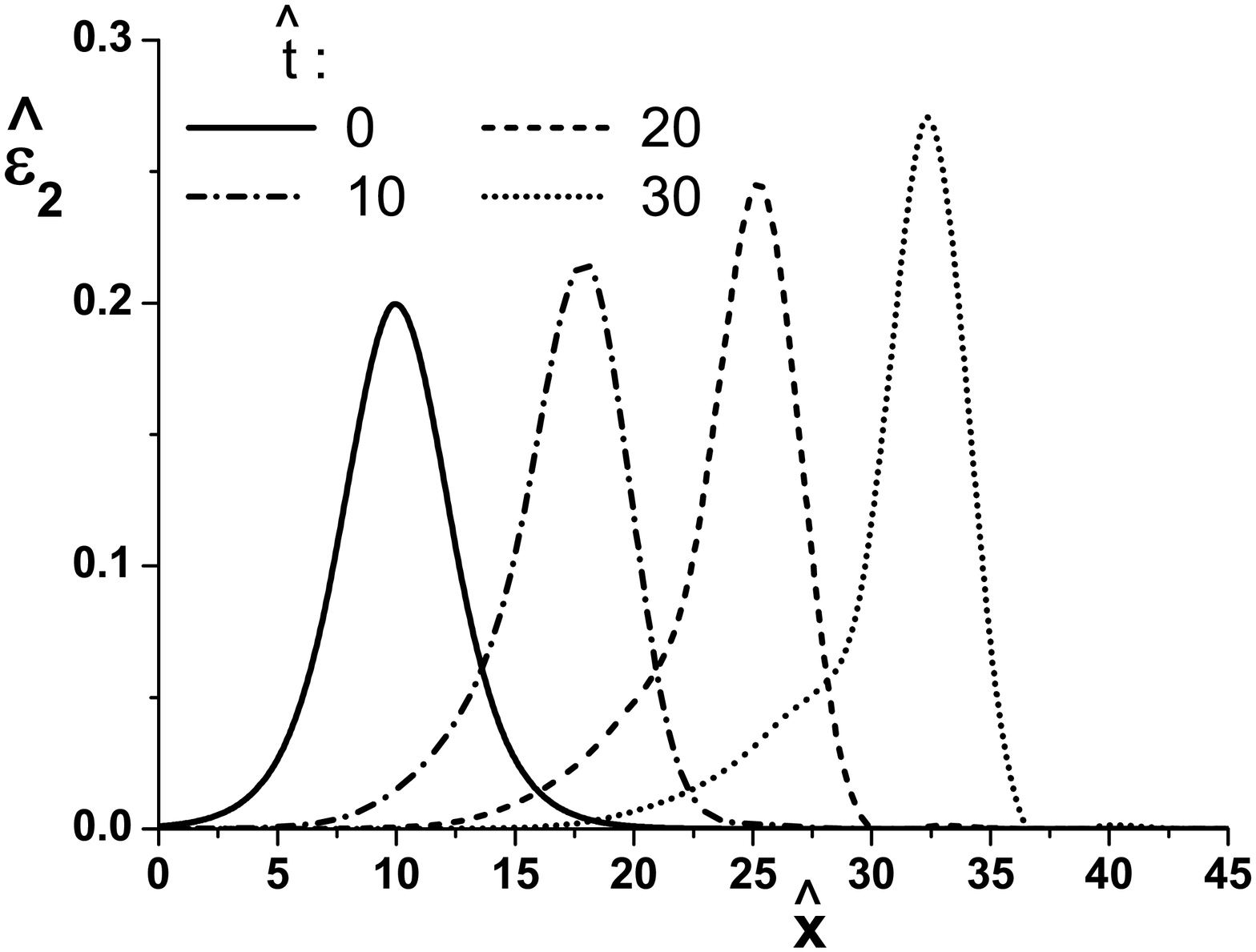}}\\
\subfigure[ ]{\label{fig4c}
\includegraphics[width=0.485\textwidth]{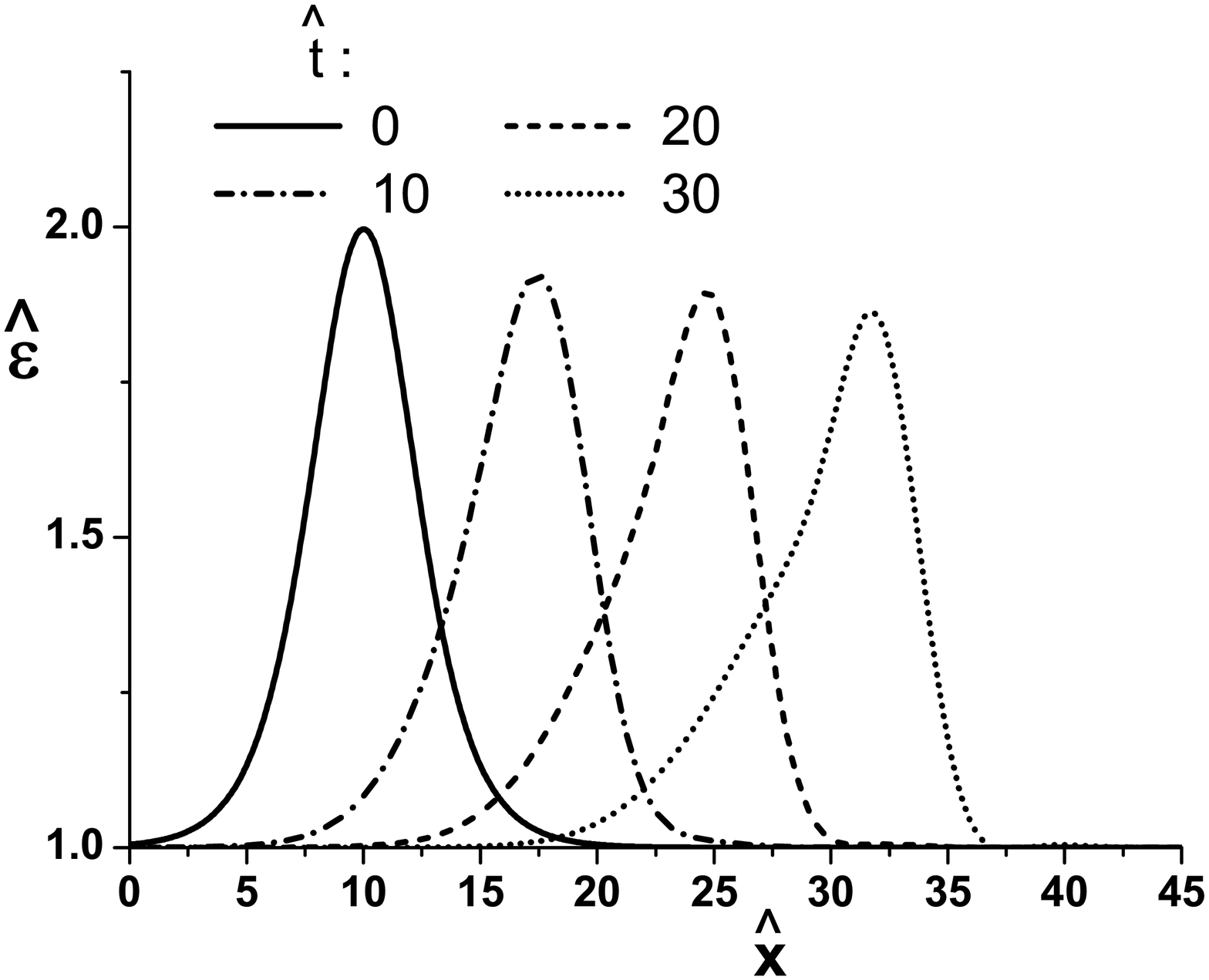}}
\end{center}
\caption{Numerical solutions for the energy density disturbances in the nonlinear regime in Eqs.\ (29) (fig. \ref{fig4a}) and (30) (fig. \ref{fig4b}) for $\eta_0/s_{0}=1/(4\pi)$ and $\hat{\tau}_{\pi}=[2-ln(2)]/(2\pi)$.  The initial conditions are (\ref{sle1}) and (\ref{sle2}) with $A_{1}=0.8$, $A_{2}=0.2$ and $B_{1}=B_{2}=3$ .  In this case the width of the initial pulses is 6 times larger than in Fig.\ \ref{fig2}. The fig. \ref{fig4c} shows the complete energy density perturbation.  The perturbations with these initial profiles mimic
soliton behavior.}
\label{fig4}
\end{figure}

The calculations shown in Fig.\ \ref{fig2} are repeated in Fig.\ \ref{fig4} (same transport coefficients) now with different initial conditions, i.e, larger widths $B_{1}=B_{2}=3$ .
Fig.\ \ref{fig4} shows an intermediate configuration between shock wave formation (wall formation) and an approximately stable soliton propagation for the total perturbation $\hat{\varepsilon}$ in \ref{fig4c}.  We note that pulses with larger width are not only more stable but the second order effects become more significant for larger times.

\begin{figure}[ht!]
\begin{center}
\subfigure[ ]{\label{fig5a}
\includegraphics[width=0.485\textwidth]{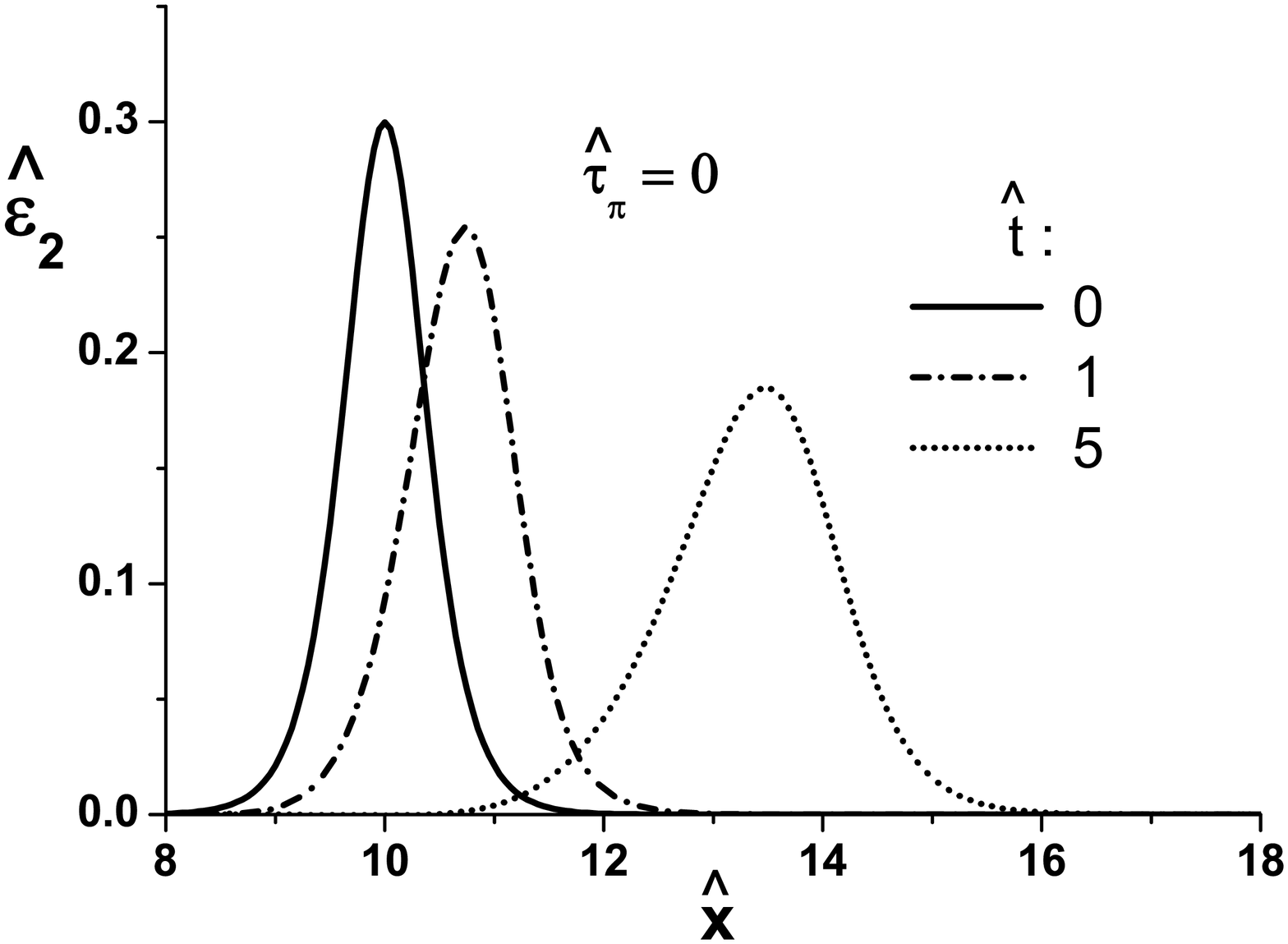}}
\subfigure[ ]{\label{fig5b}
\includegraphics[width=0.485\textwidth]{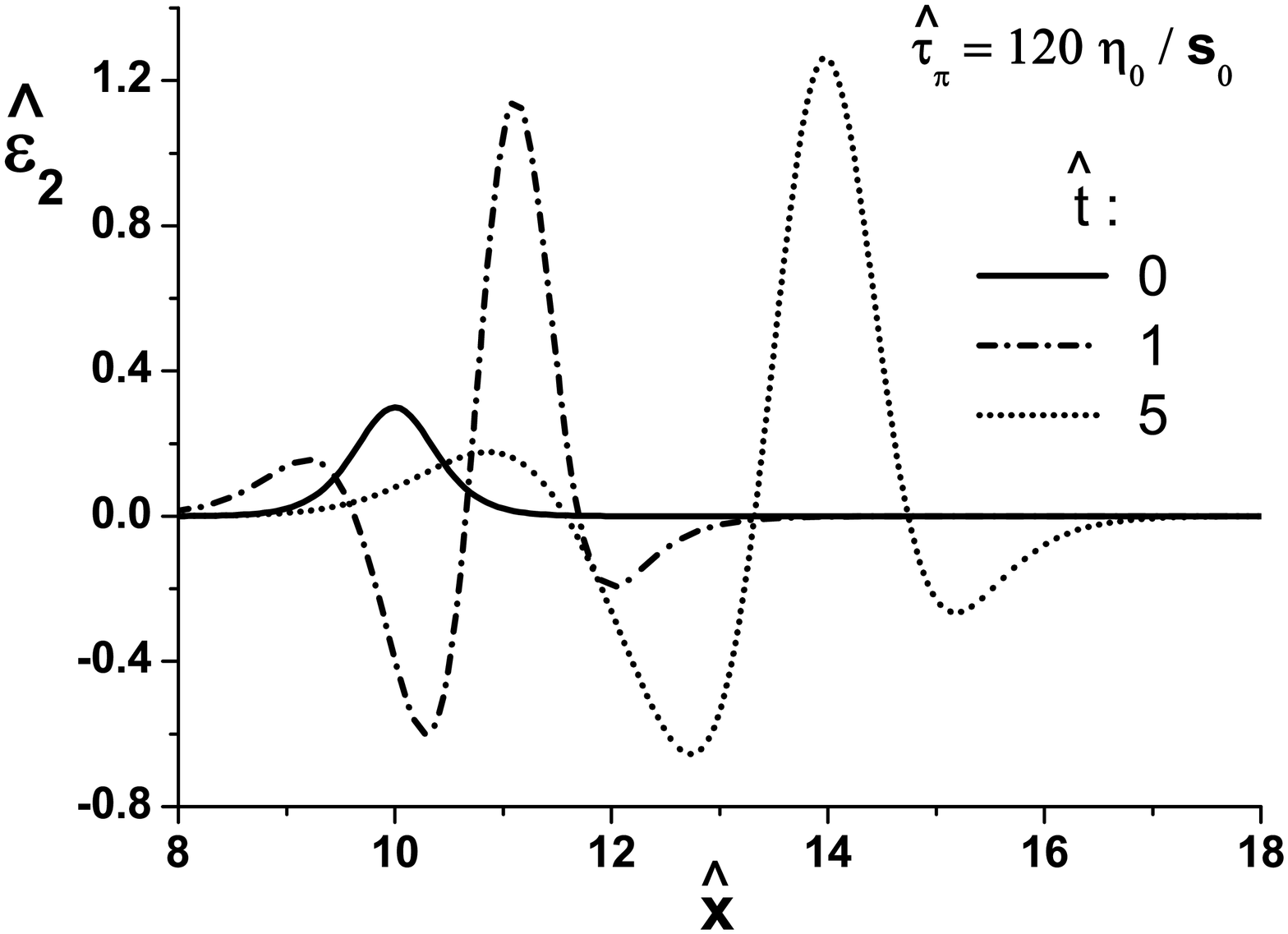}}\\
\subfigure[ ]{\label{fig5c}
\includegraphics[width=0.485\textwidth]{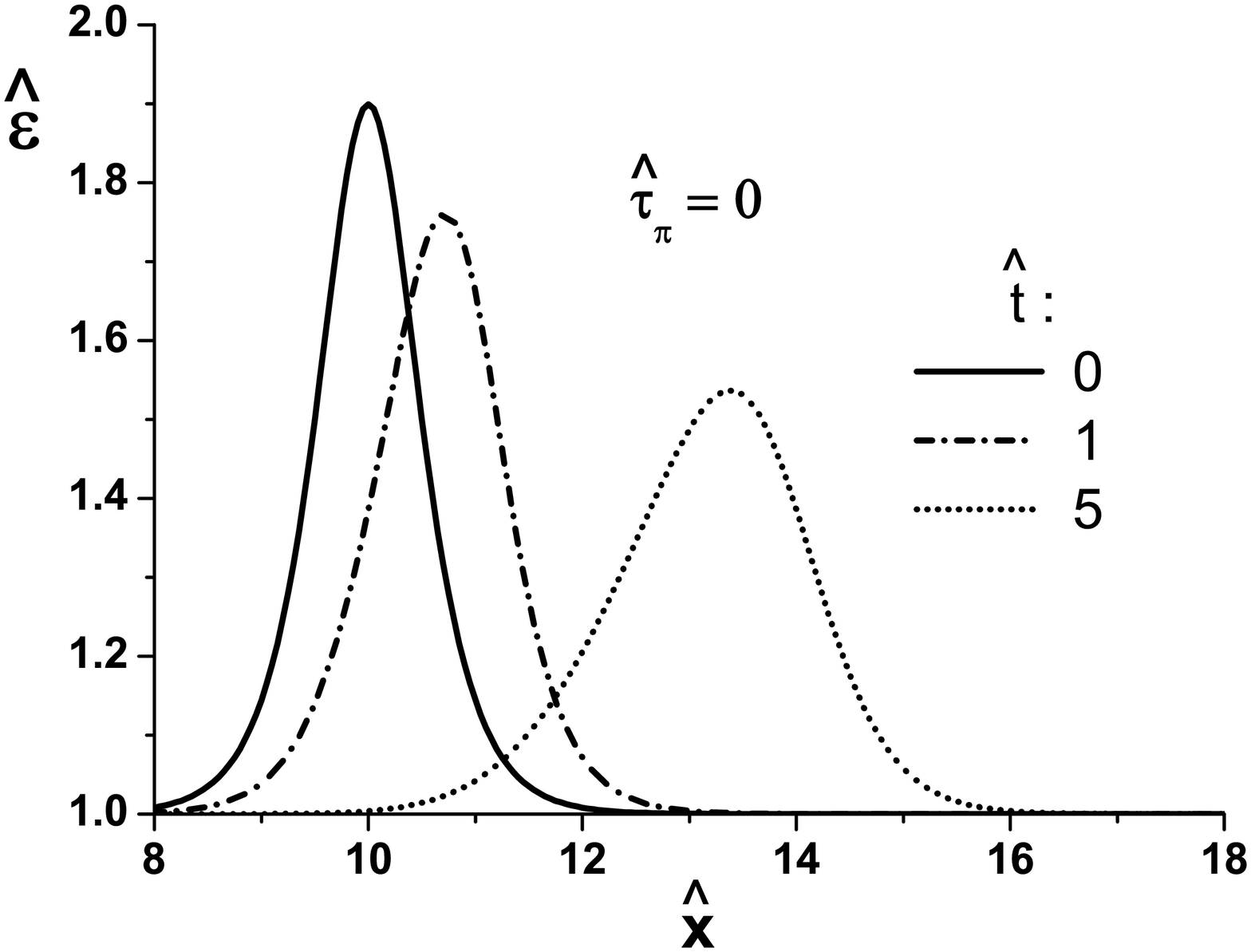}}
\subfigure[ ]{\label{fig5d}
\includegraphics[width=0.485\textwidth]{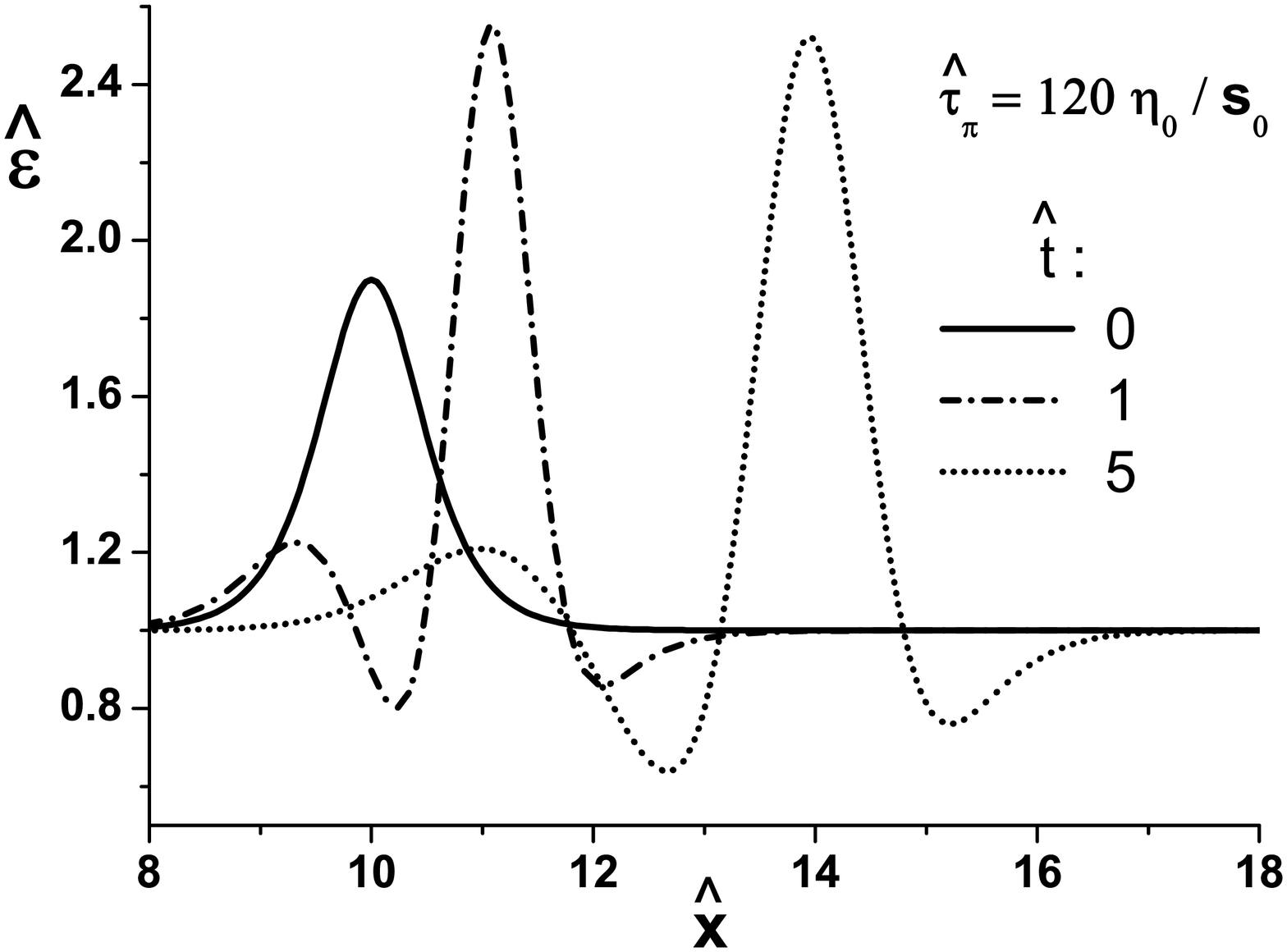}}
\end{center}
\caption{Numerical solutions for the energy density disturbance in the nonlinear regime in Eq.\ (30) {(fig. \ref{fig5a}}  and fig. \ref{fig5b}) for $\eta_0/s_{0}=1/(4\pi)$ and two choices of $\hat{\tau}_\pi$.  The initial conditions are (\ref{sle1}) and (\ref{sle2}) with $A_{1}=0.6$, $A_{2}=0.3$, $B_{1}=0.7$ and $B_{2}=0.5$ .  The fig. \ref{fig5c} and \ref{fig5d} shows the complete energy density perturbation.  For large values of the relaxation time coefficient, the energy perturbation $\hat{\varepsilon}_{2}$ acquires large amplitude and becomes inconsistent as a small disturbance (note, however, that the initial gradients are large).}
\label{fig5}
\end{figure}

In Fig.\ \ref{fig5} we considered $A_{1}=0.6$, $A_{2}=0.3$, $B_{1}=0.7$, and $B_{2}=0.5$ for a small viscosity $\eta_0/s_{0}=1/(4\pi)$ and varied the value of the relaxation time coefficient.  The values considered were $\hat{\tau}_{\pi}=0$ (Navier-Stokes limit) and
$\hat{\tau}_{\pi}=120 \eta_0/s_0$ (where the relaxation time is much more important than the shear viscosity). We only plot the perturbations affected by relaxation: $\hat{\varepsilon}_{2}$ and consequently $\hat{\varepsilon}$. We notice that the size of the perturbations increase when one increases the relaxation coefficient. This limit is not a very plausible choice but it is interesting to see that the resulting solutions are unstable since they generate values
of $\hat{\varepsilon}_{2}$ which are unacceptably large. If we consider that pulses originate from inhomogeneous density profiles or quantum fluctuations, it is reasonable to assume that the most realistic pulses could be in principle more localized in space and thus they would suffer dissipative, nonlinear, and dispersive effects losing its localized profile.

In Eq.\ (\ref{weeps2onlyad}) the terms with  $\eta_0/s_{0}$ (except for the last one) contribute to dissipation. The last term of (\ref{weeps2onlyad})
introduces dispersion and involves both $\eta_0/s_{0}$ and the combination $ \Delta_3 = \eta_0/s_{0}-\hat{\tau}_{\pi}$.  When $\hat{\tau}_{\pi}$ tends to zero we recover the Navier-Stokes limit, where problems with causality and instability are expected to appear. When $\hat{\tau}_{\pi}$ becomes very large, in principle, no problem was expected to occur. However the very large amplification of the amplitude $\hat{\varepsilon}_{2}$ is surprising.
It implies that a large amount of energy is transferred from the medium to the wave. We see here evidence that the large value chosen for
$\hat{\tau}_{\pi}$ in this particular configuration may be unphysical. This is a interesting finding since in the linear perturbative limit (discussed before in Section II) there were no apparent inconsistencies associated with large values of $\hat{\tau}_{\pi}$. The existence of an upper bound for $\hat{\tau}_{\pi}$ can only be seen in the nonlinear perturbation theory used here. However, one may also interpret this enhancement in the amplitude as an indication that the higher order terms that were neglected in the expansion have become significant and must be taken into account (the initial profile is such that the initial spatial gradients are not very small). It would be interesting to check if this nonlinear instability can appear in the existing numerical hydrodynamic codes.

\begin{figure}[ht!]
\begin{center}
\subfigure[ ]{\label{fig6a}
\includegraphics[width=0.485\textwidth]{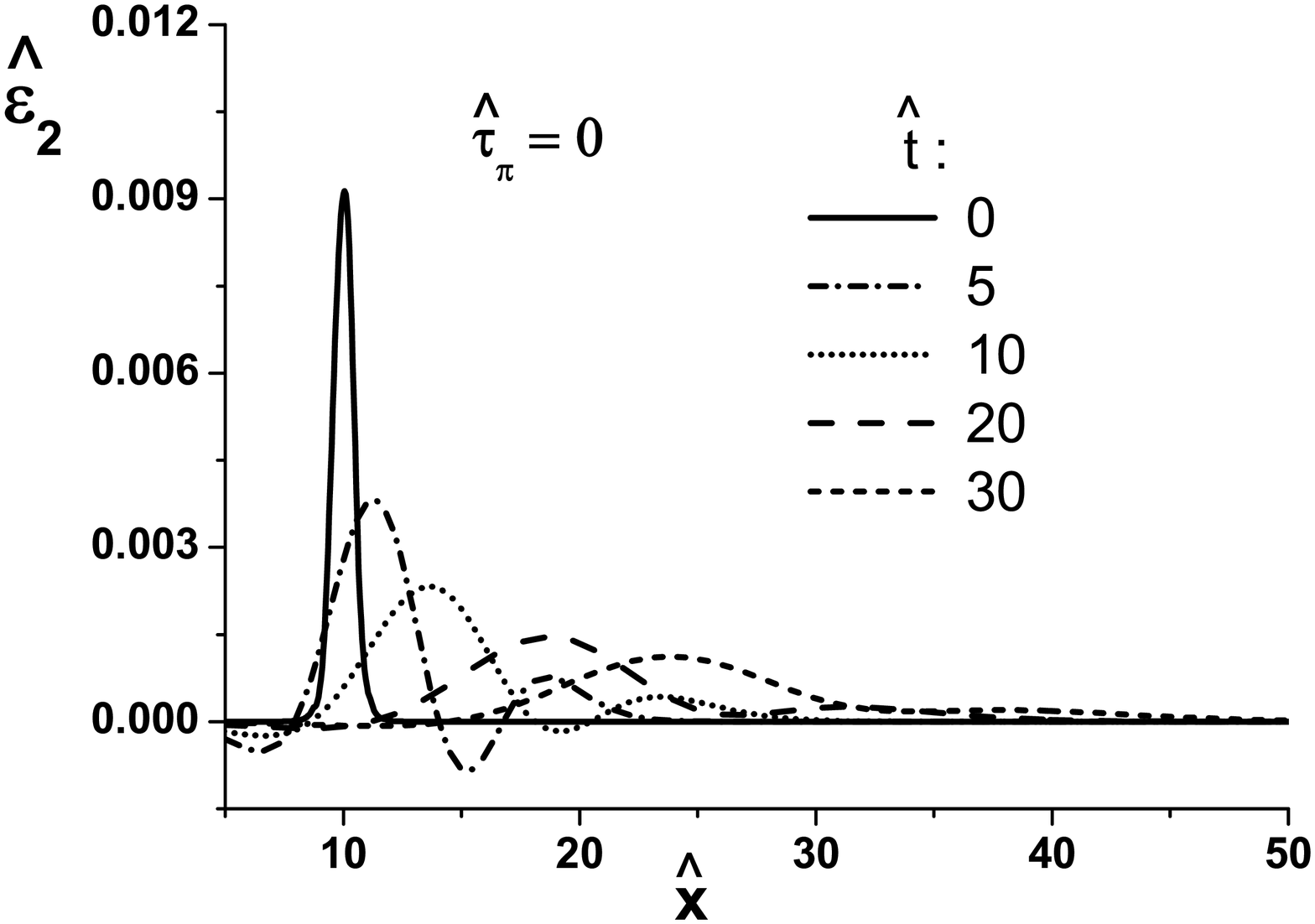}}
\subfigure[ ]{\label{fig6b}
\includegraphics[width=0.485\textwidth]{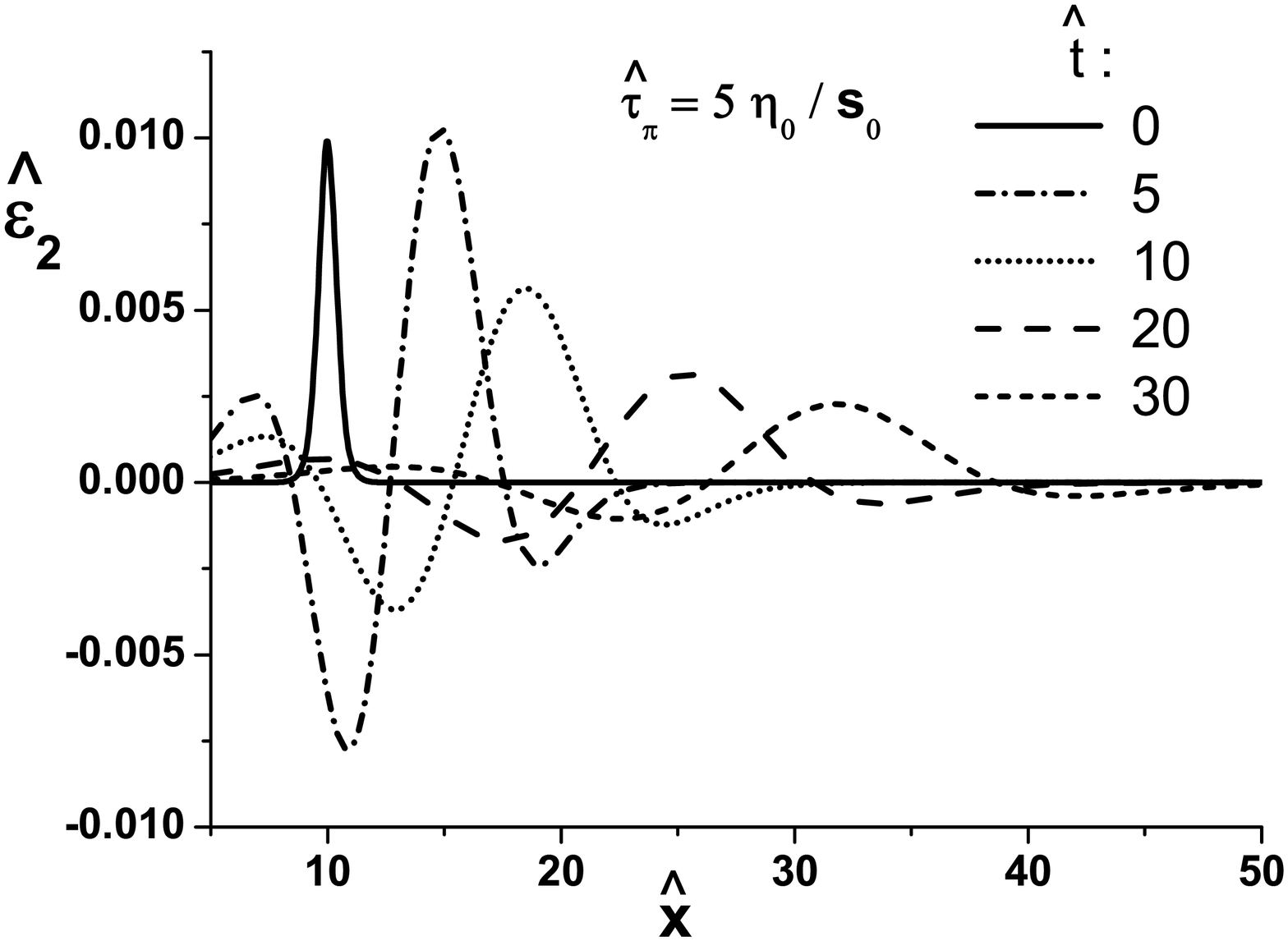}}\\
\subfigure[ ]{\label{fig6c}
\includegraphics[width=0.485\textwidth]{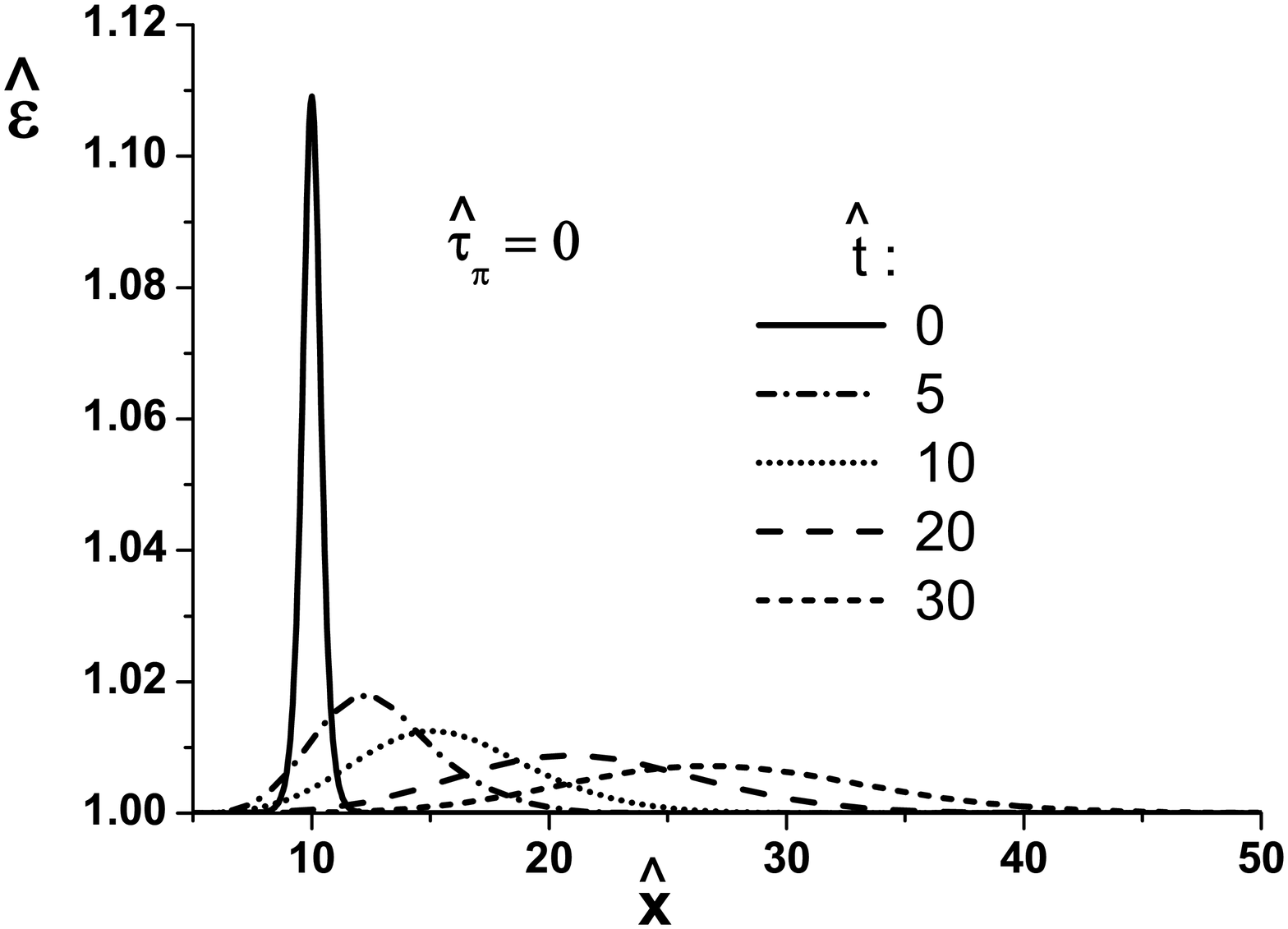}}
\subfigure[ ]{\label{fig6d}
\includegraphics[width=0.485\textwidth]{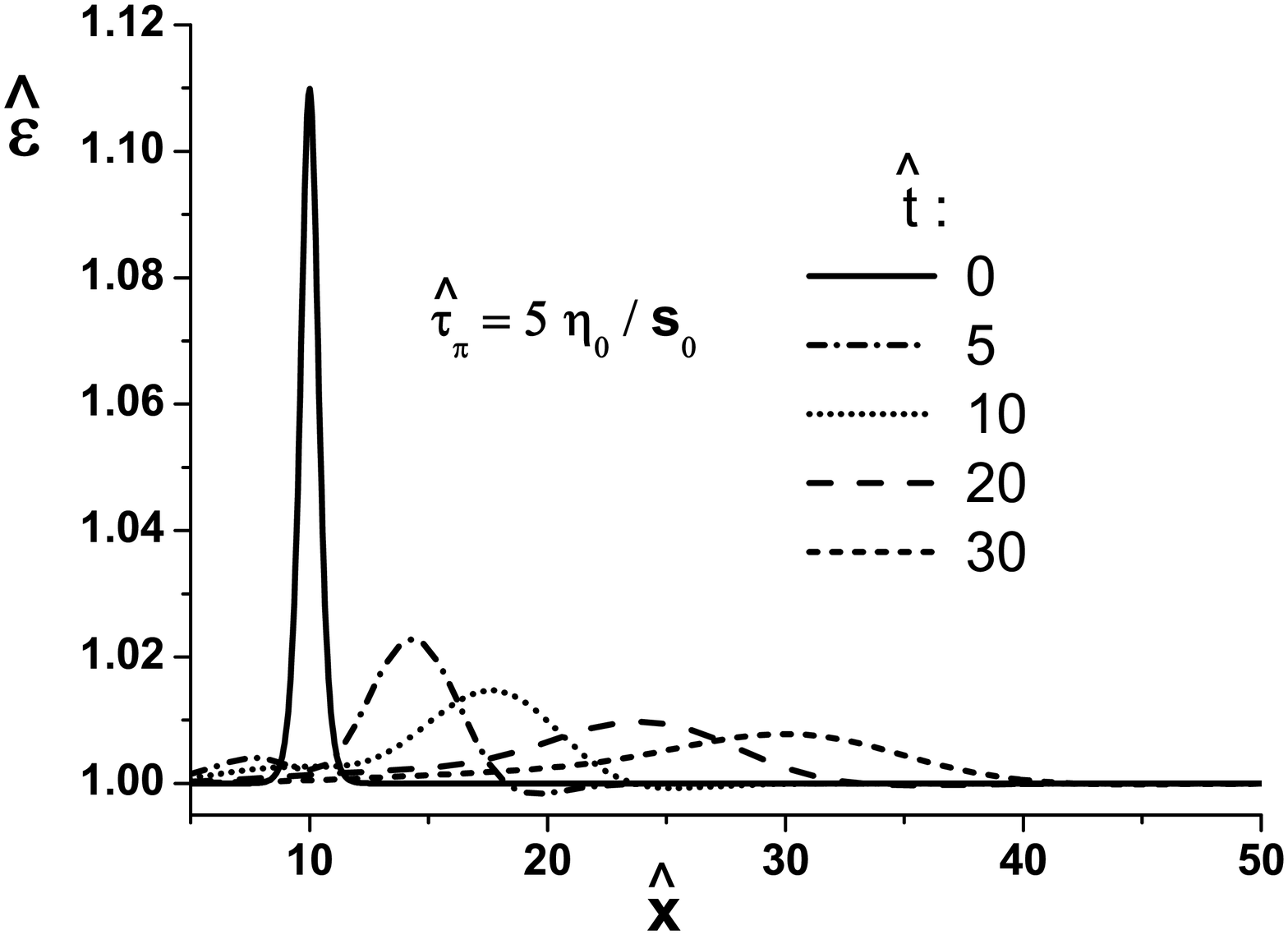}}
\end{center}
\caption{Comparison between the energy density perturbations in Navier-Stokes and Israel-Stewart theory for $\eta_0/s_{0}=1$ and $\hat{\tau}_\pi=5\,\eta_0/s_0$.  The initial conditions are (\ref{sle1}) and (\ref{sle2}) with $A_{1}=0.1$, $A_{2}=0.01$ and $B_{1}=B_{2}=0.5$ .  The fig. \ref{fig6a}  and fig. \ref{fig6b} show the numerical solutions for Eq.\ (30) and
fig. \ref{fig6c} and \ref{fig6d} are the complete energy density perturbation.  The relaxation
ensures that rarefaction occurs in the tail of the pulse while there is an enhancement in the front of the pulse. }
\label{fig6}
\end{figure}

In Fig.\ \ref{fig6} for a large viscosity $\eta_0/s_{0}=1$ and small amplitudes and widths, given by $A_{1}=0.1$, $A_{2}=0.01$ and $B_{1}=B_{2}=0.5$,  we compare the results for two different theories, NS and IS. In this case $\hat{\tau}_{\pi}=0$ (Navier-Stokes case) and
$\hat{\tau}_{\pi}=5 \eta_0/s_0$, which is a reasonable estimate for $\hat{\tau}_{\pi}$ for systems described by the Boltzmann equation. This figure is analogous to Fig.\ \ref{fig5}, as the Israel-Stewart fluid ensures that rarefaction occurs in the tail and there is an enhancement in the front of the pulse.

\begin{figure}[ht!]
\begin{center}
\subfigure[ ]{\label{fig7a}
\includegraphics[width=0.485\textwidth]{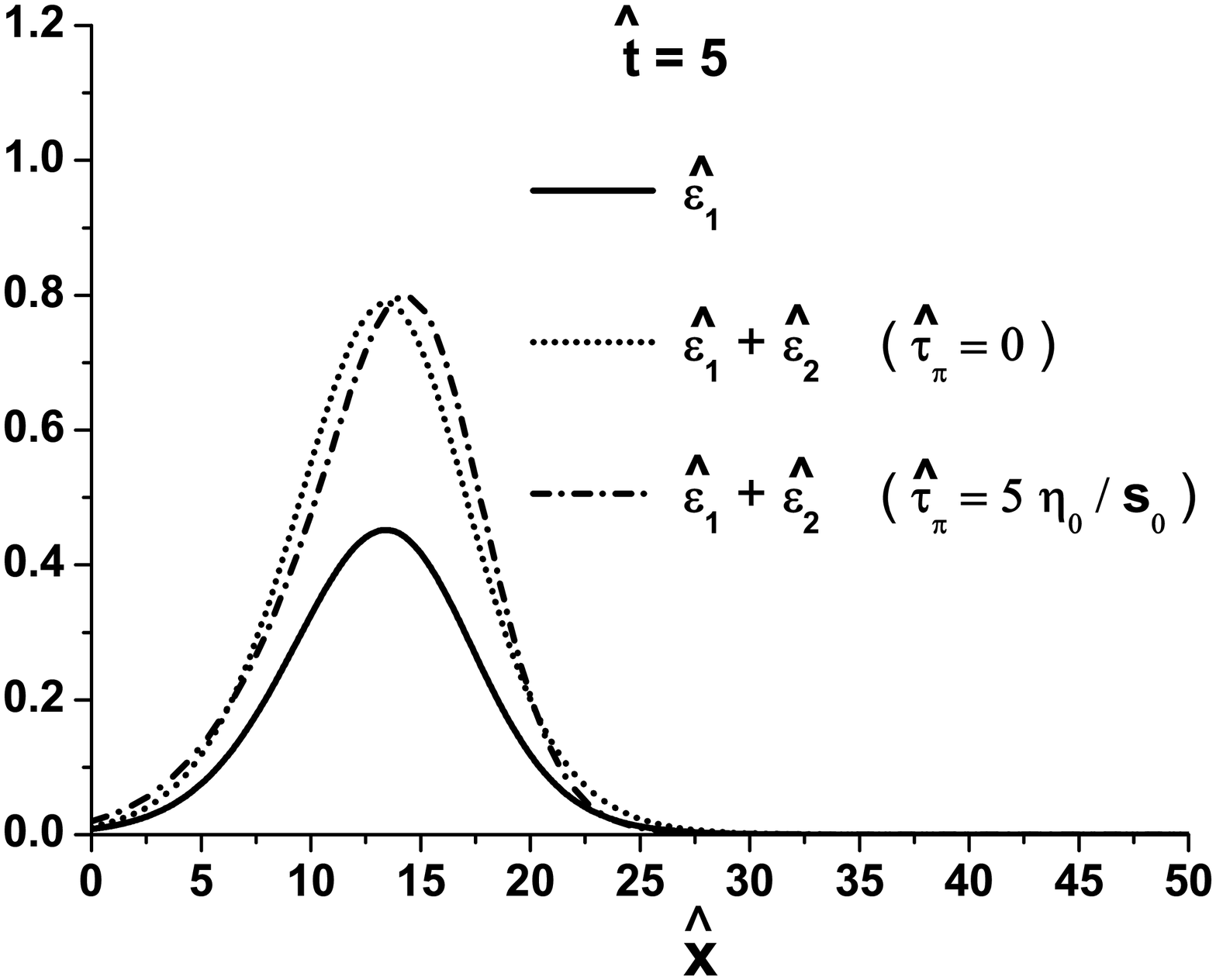}}
\subfigure[ ]{\label{fig7b}
\includegraphics[width=0.485\textwidth]{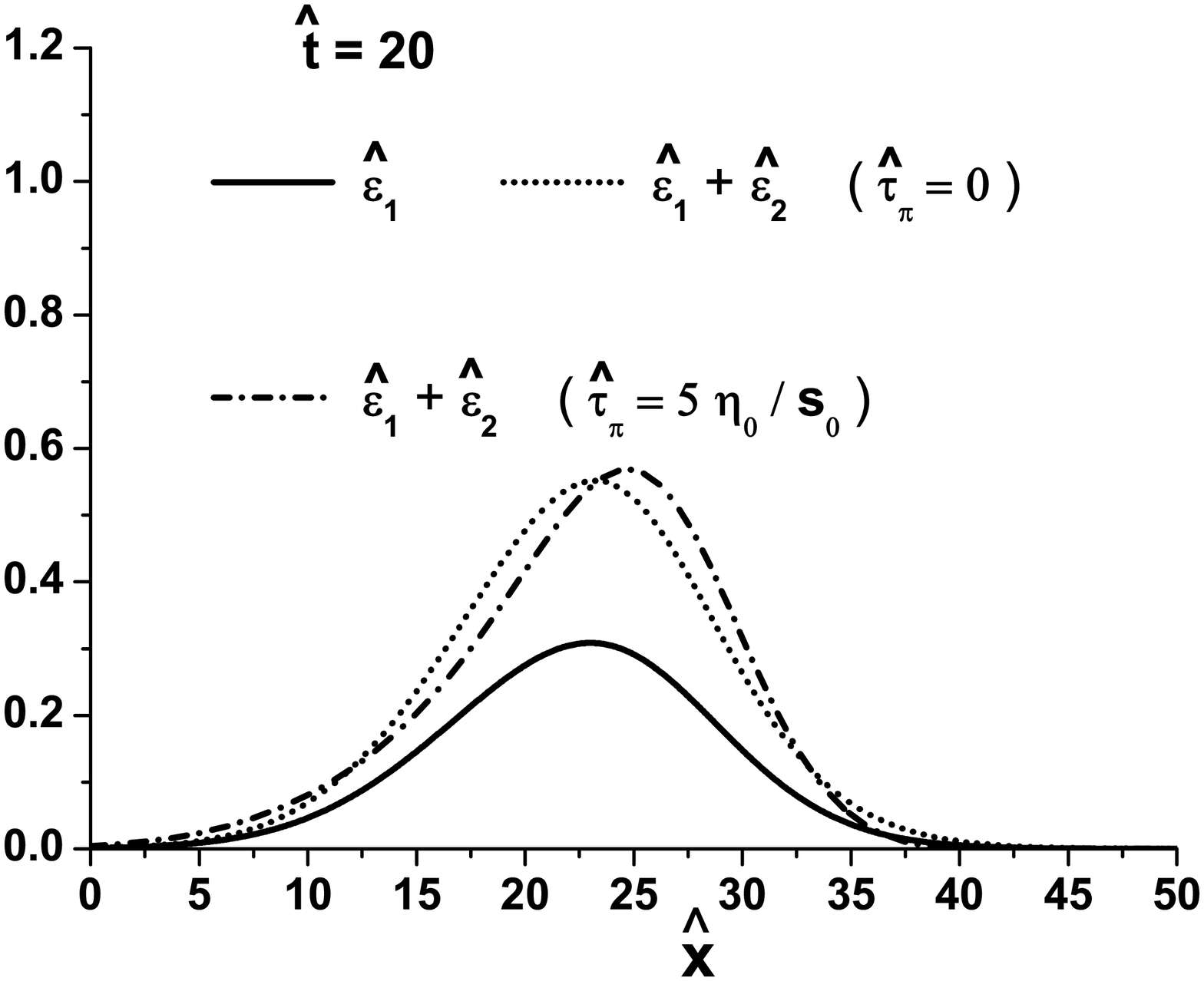}}
\end{center}
\caption{Comparison between the energy density perturbations in Navier-Stokes and Israel-Stewart theory for $\eta_0/s_{0}=1$ and $\hat{\tau}_\pi=5\,\eta_0/s_0$ for larger width of the initial pulse.  The initial conditions are (\ref{sle1}) and (\ref{sle2}) with $A_{1}=0.6$, $A_{2}=0.4$ and $B_{1}=B_{2}=4$ .  Conformal Israel-Stewart hydrodynamics favors the wall front formation.}
\label{fig7}
\end{figure}

Using the same parameters as in Fig.\ \ref{fig6}, we summarize the effects of relaxation considering the ``soliton-like'' configuration for the initial conditions: $A_{1}=0.6$, $A_{2}=0.4$, $B_{1}=B_{2}=4$ in Fig.\ \ref{fig7}.  Relaxation increases the pulse amplitudes in some regions, as it has a dispersive character. However, this behavior is different from the NS case in which there is an enhancement of the amplitude in the opposite direction of the pulse.

The pulse in the Israel-Stewart fluid propagates ahead of that from the Navier-Stokes fluid. We clearly notice that IS hydrodynamics favors the wall front formation, while NS disperses the pulse to the opposite direction of motion. This might be the most important feature of relaxation time effects in nonlinear wave perturbation found in this paper and is both present for strong and weak coupling inspired parameters.

\subsection{Gaussian initial profile}

Again, we consider the strong coupling parameters $3\chi/4=\eta_0/s_{0}=1/(4\pi)$ and two values for $\hat{\tau}$.
We solve (\ref{burgXTcad}) with the following gaussian initial condition:
\begin{equation}
\hat{\varepsilon}_{1}(\hat{x},0)= C_{1} \ e^{-(\hat{x}/D_{1})^{2}}
\label{gle1}
\end{equation}
and insert the obtained numerical solution of (\ref{burgXTcad}) into (\ref{weeps2onlyad}) with the initial gaussian profile for
$\hat{\varepsilon}_{2}$:
\begin{equation}
\hat{\varepsilon}_{2}(\hat{x},0)= C_{2} \ e^{-(\hat{x}/D_{2})^{2}}.
\label{gle2}
\end{equation}
The amplitudes $C_{1},\,C_{2}$ and the widths $D_{1},\,D_{2}$ are chosen to study some stability features.

\begin{figure}[ht!]
\begin{center}
\subfigure[ ]{\label{fig8a}
\includegraphics[width=0.485\textwidth]{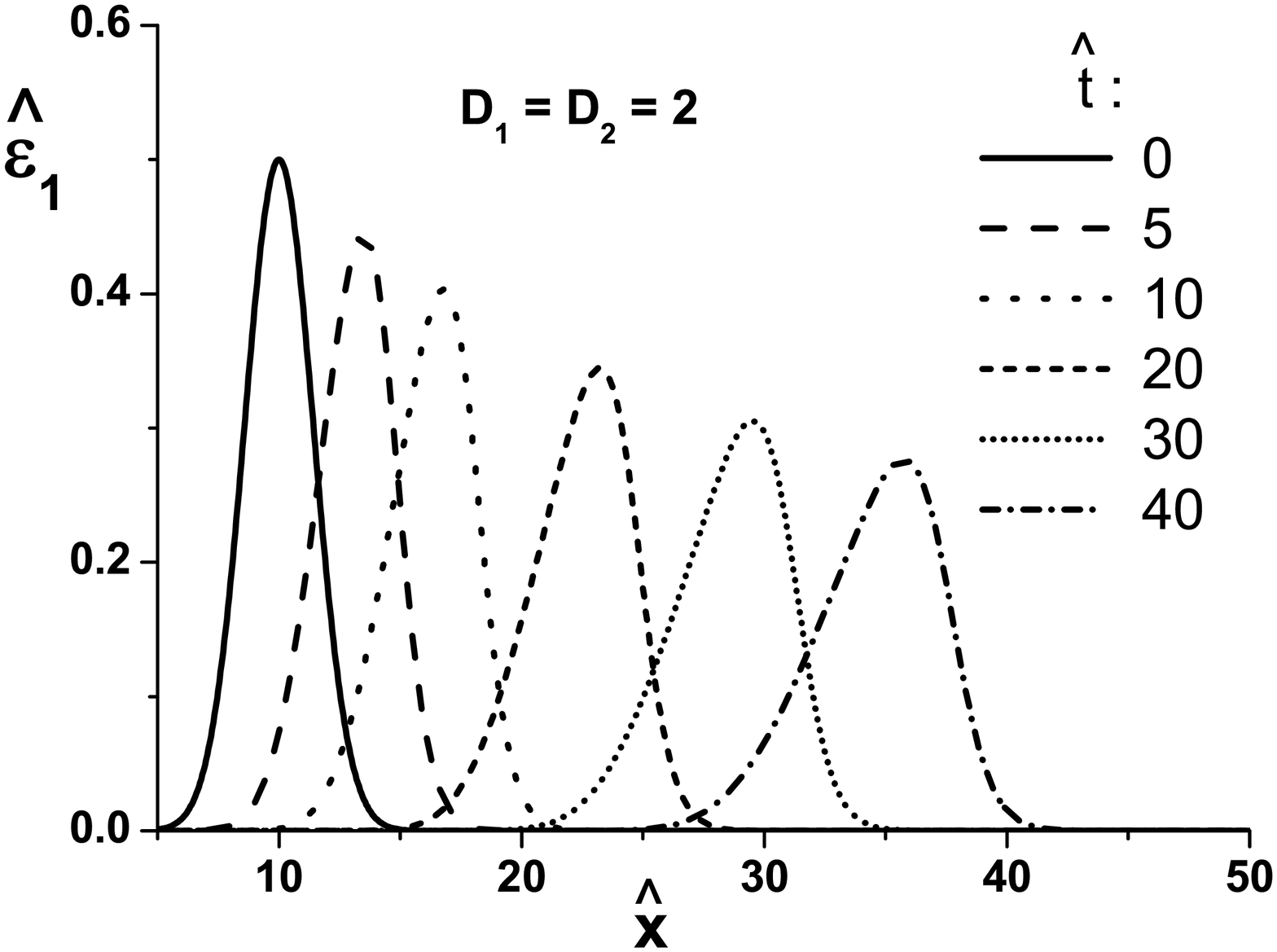}}
\subfigure[ ]{\label{fig8b}
\includegraphics[width=0.485\textwidth]{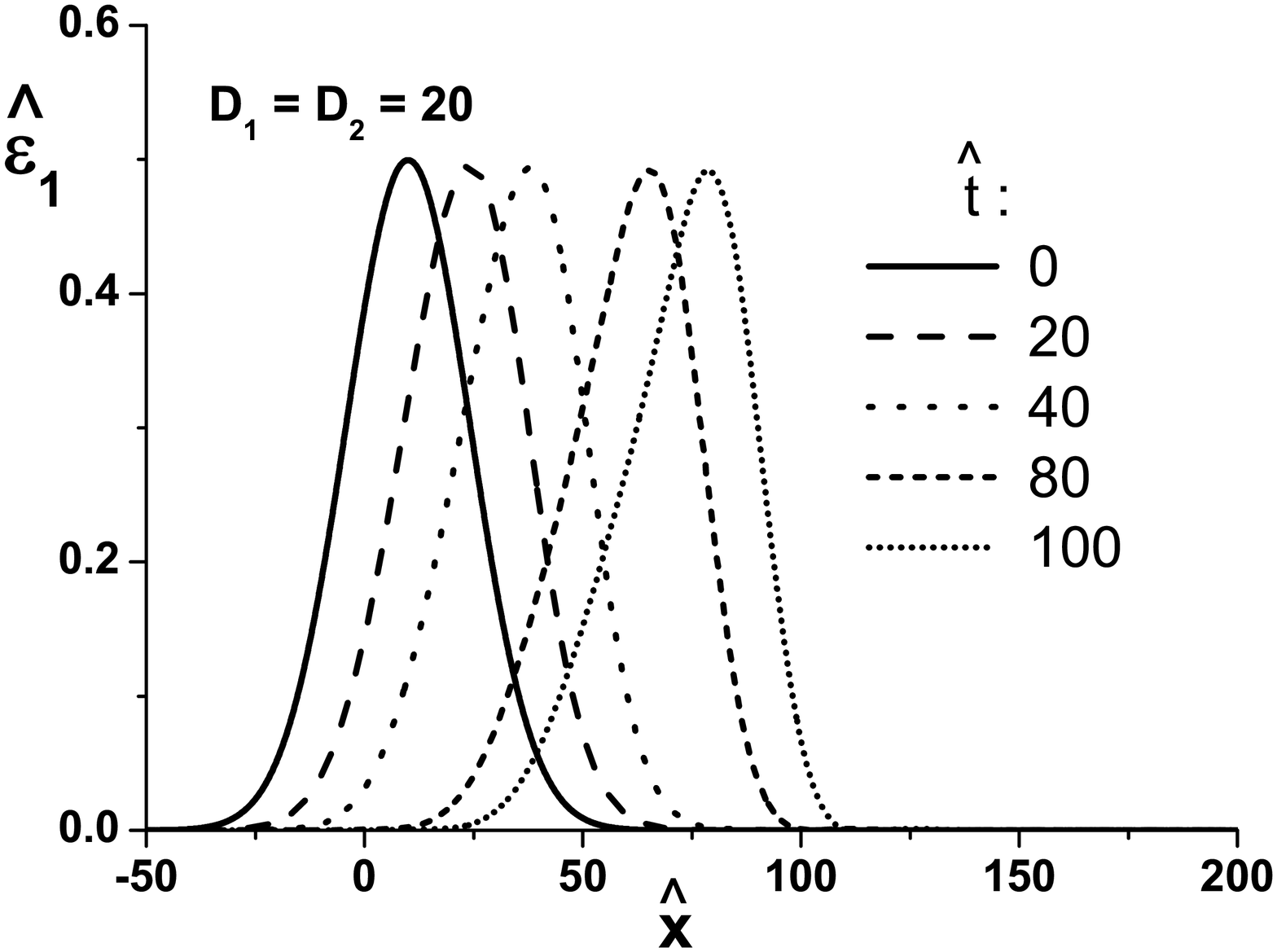}}\\
\subfigure[ ]{\label{fig8c}
\includegraphics[width=0.485\textwidth]{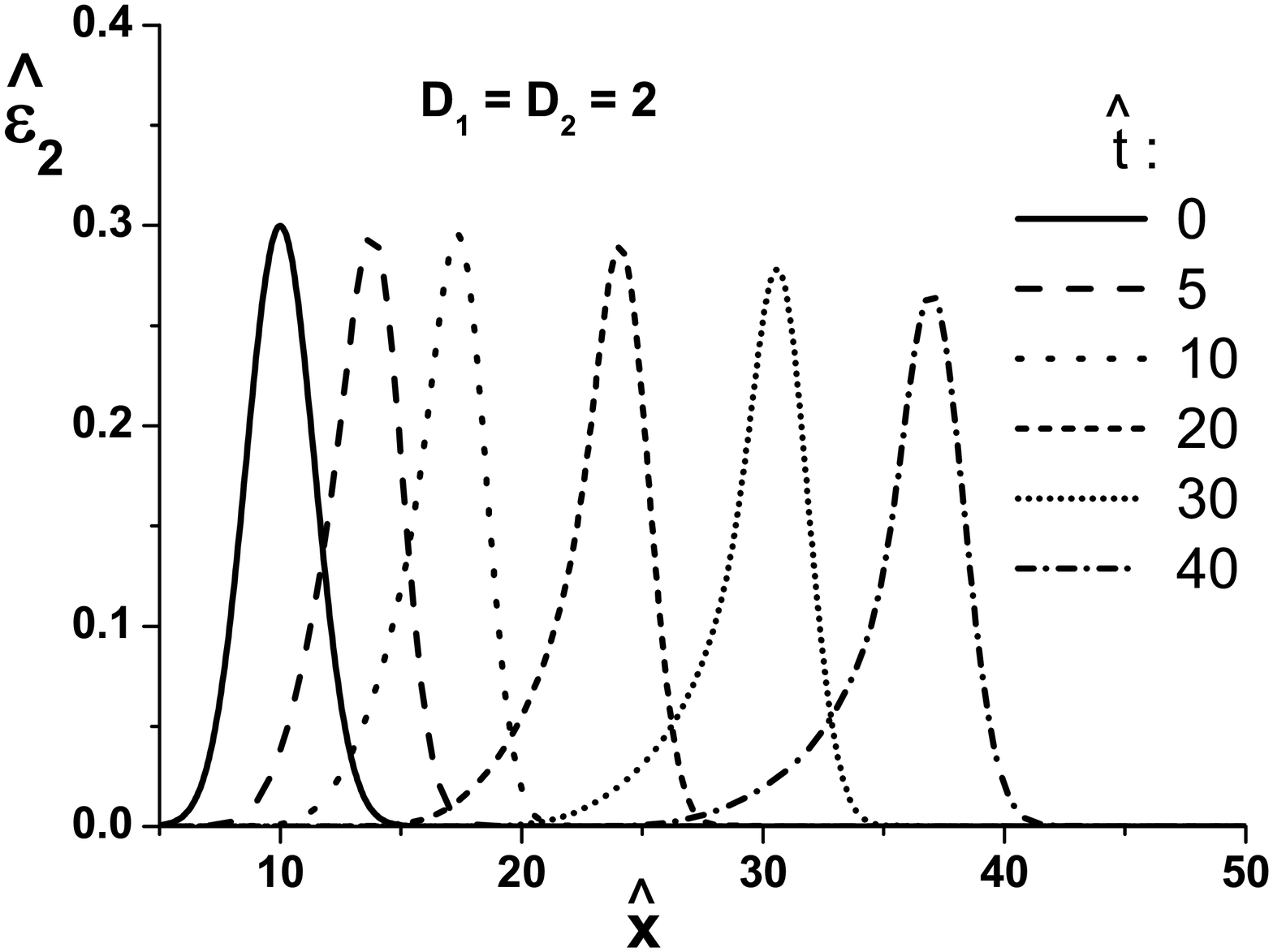}}
\subfigure[ ]{\label{fig8d}
\includegraphics[width=0.485\textwidth]{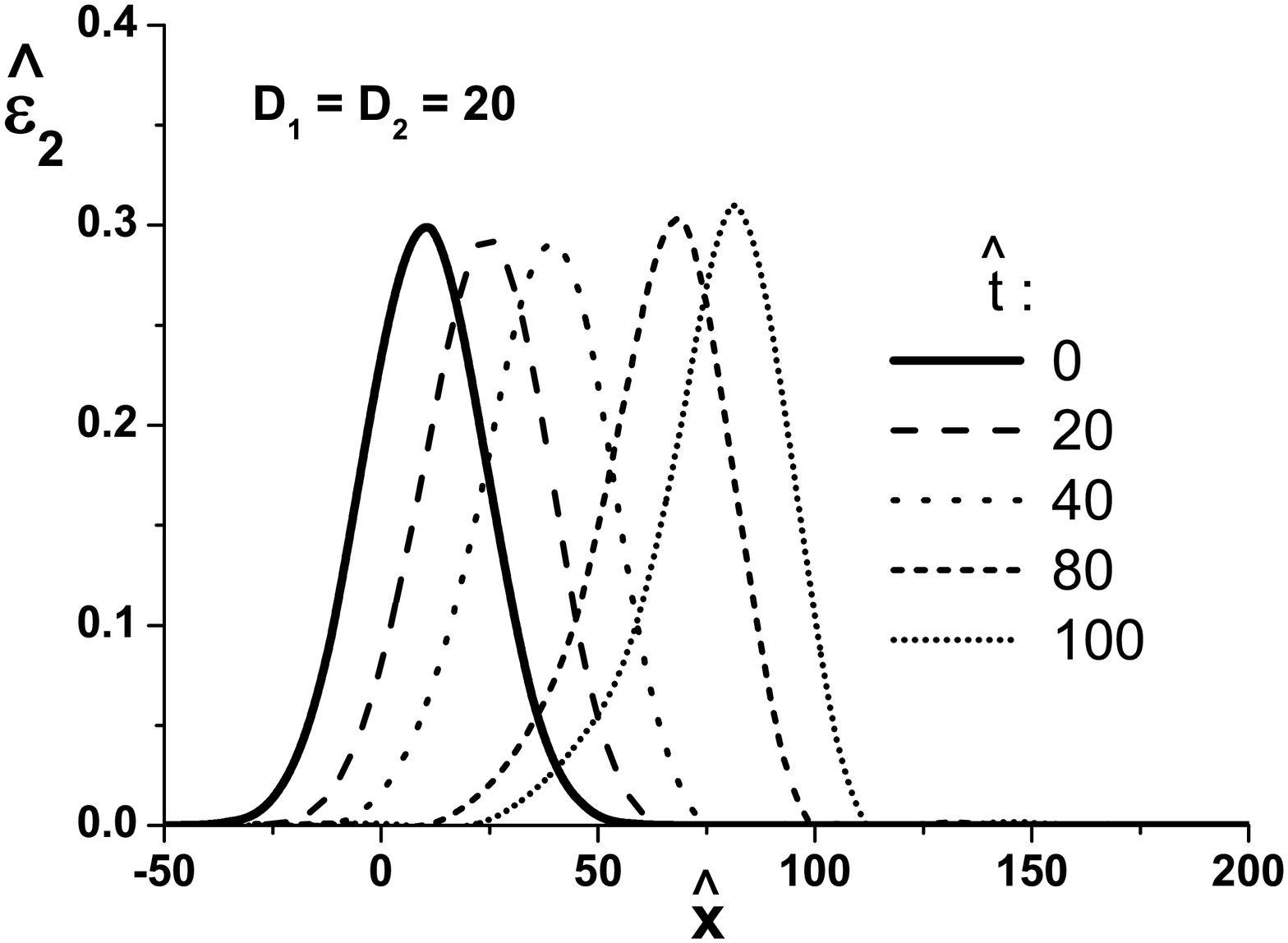}}\\
\subfigure[ ]{\label{fig8e}
\includegraphics[width=0.485\textwidth]{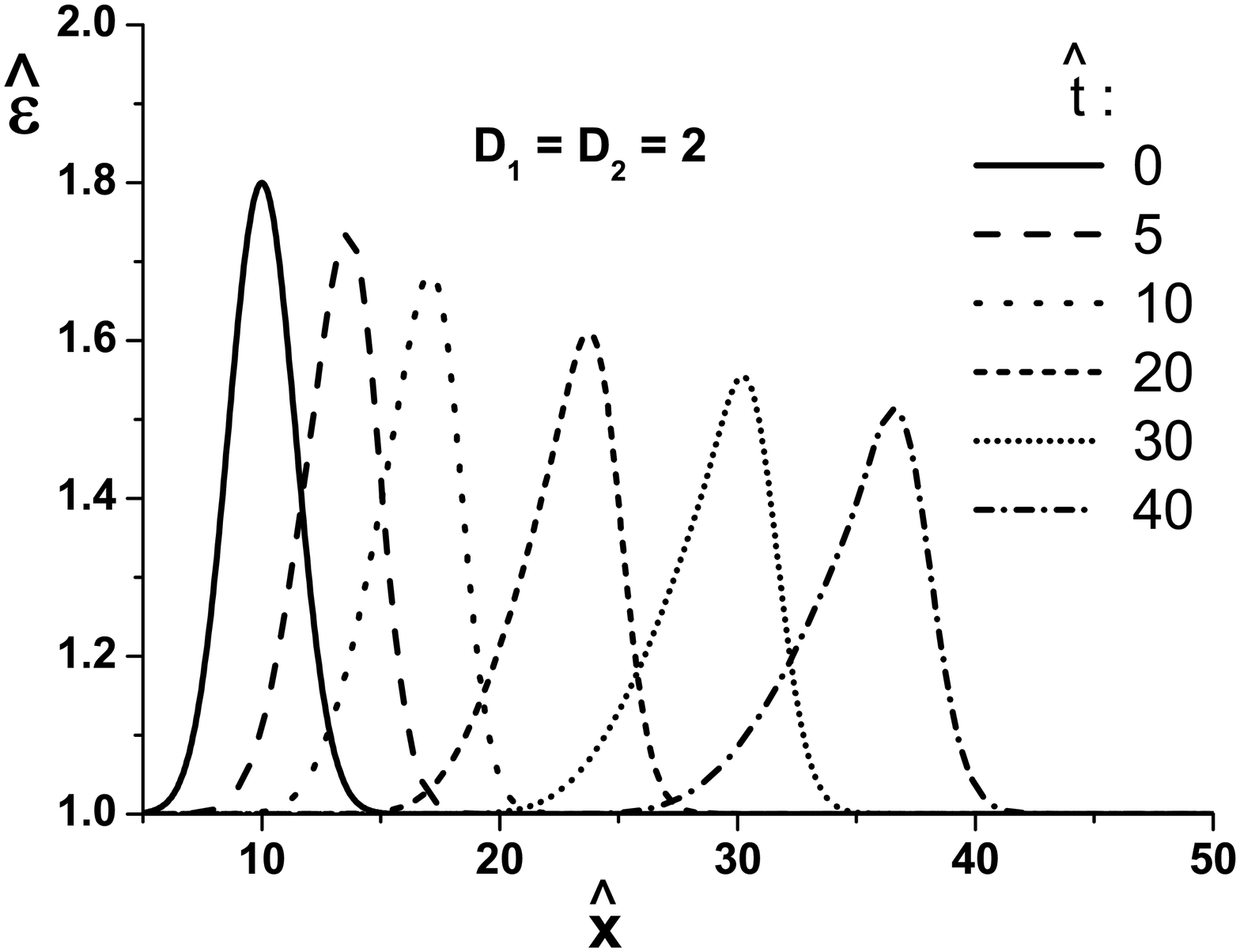}}
\subfigure[ ]{\label{fig8f}
\includegraphics[width=0.485\textwidth]{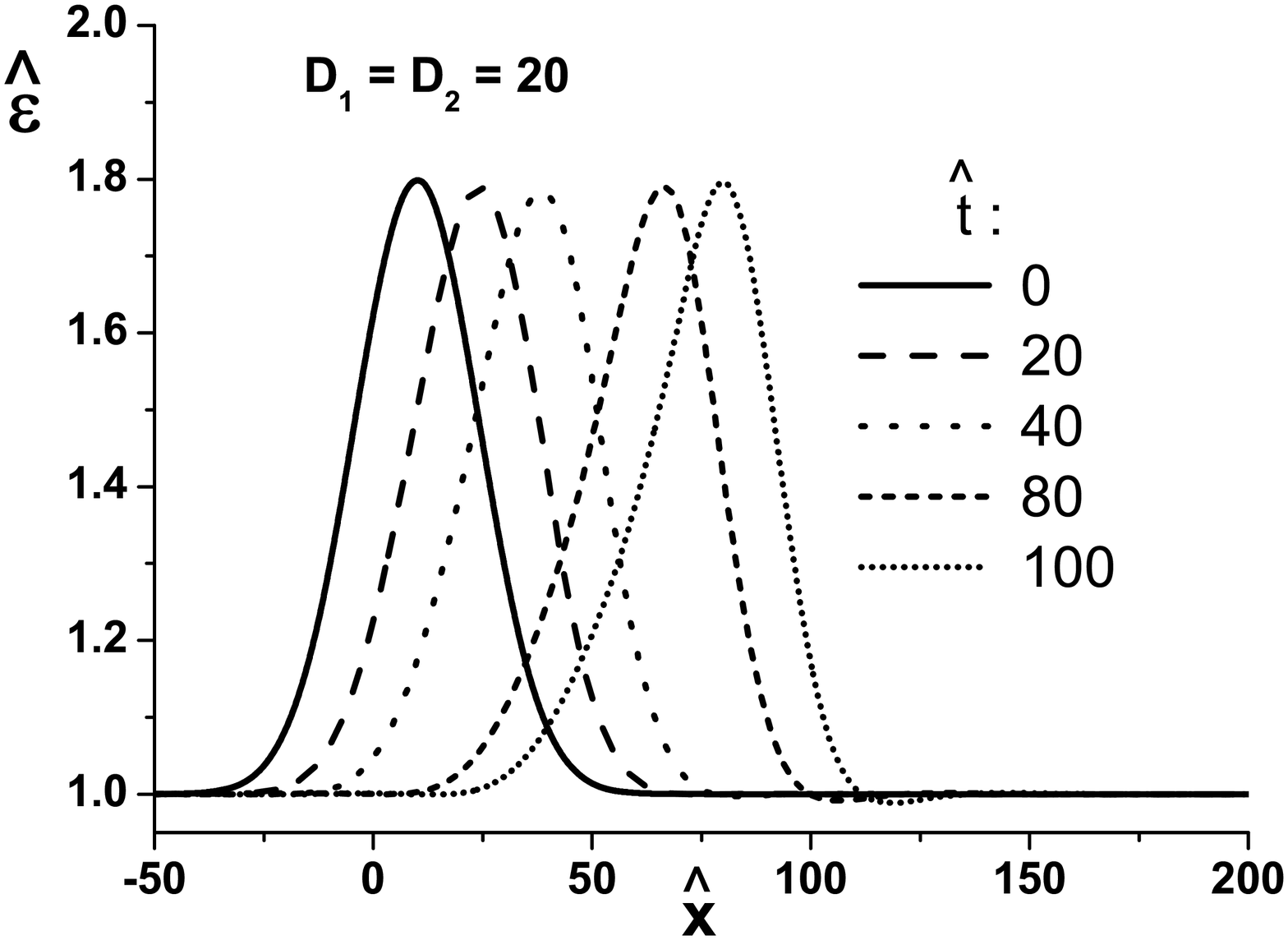}}
\end{center}
\caption{Stability is found by increasing the initial width of the initial gaussian profiles
(\ref{gle1}) and (\ref{gle2}) with $C_{1}=0.5$ and $C_{2}=0.3$ . The plots are the numerical solutions in the nonlinear regime in Eqs. (29) (\ref{fig8a} and \ref{fig8b}), and (30) (\ref{fig8c} and \ref{fig8d}) for $\eta_0/s_{0}=1/(4\pi)$ and $\hat{\tau}_{\pi}=[2-ln(2)]/(2\pi)$.
In \ref{fig8e} and \ref{fig8f}: the complete energy density perturbation
$\hat\varepsilon=1+\hat{\varepsilon}_{1}+ \hat{\varepsilon}_{2}$. The perturbations survive despite the dissipative effects.}
\label{fig8}
\end{figure}

We consider $\hat{\tau}_{\pi}=[2-ln(2)]/(2\pi)$, $C_{1}=0.5$ and $C_{2}=0.3$ in Fig. \ref{fig8}. One can see that by increasing the width of the initial profile from $D_{1}=D_{2}=2$ to $D_{1}=D_{2}=20$ guarantees stability (the gradients are significantly reduced in this case). The solution of the Burgers equation (\ref{burgXTcad}) for $\hat{\varepsilon}_{1}$ mimics a soliton when $D_{1}=D_{2}=20$.

\begin{figure}[ht!]
\begin{center}
\subfigure[ ]{\label{fig9a}
\includegraphics[width=0.485\textwidth]{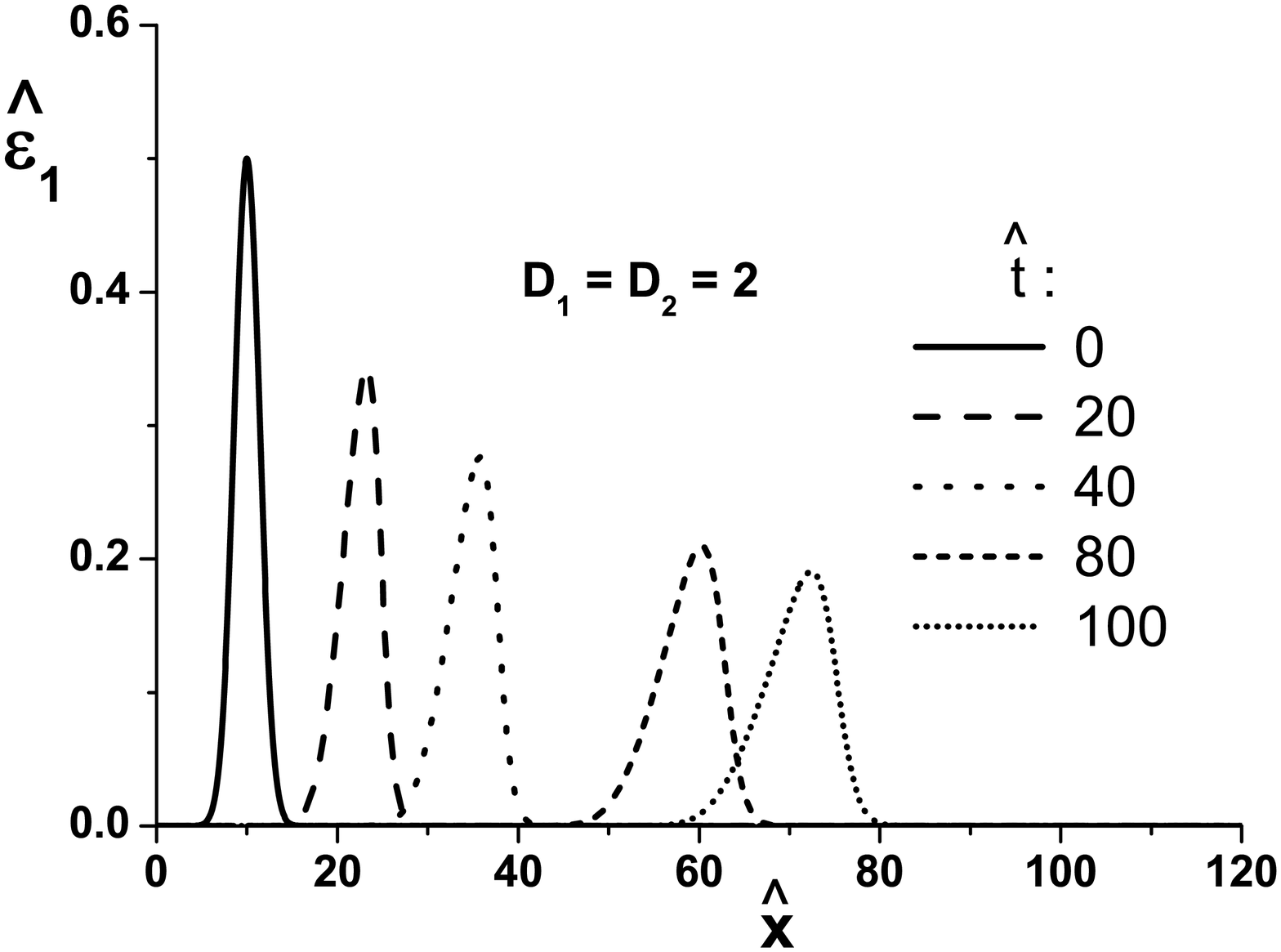}}
\subfigure[ ]{\label{fig9b}
\includegraphics[width=0.485\textwidth]{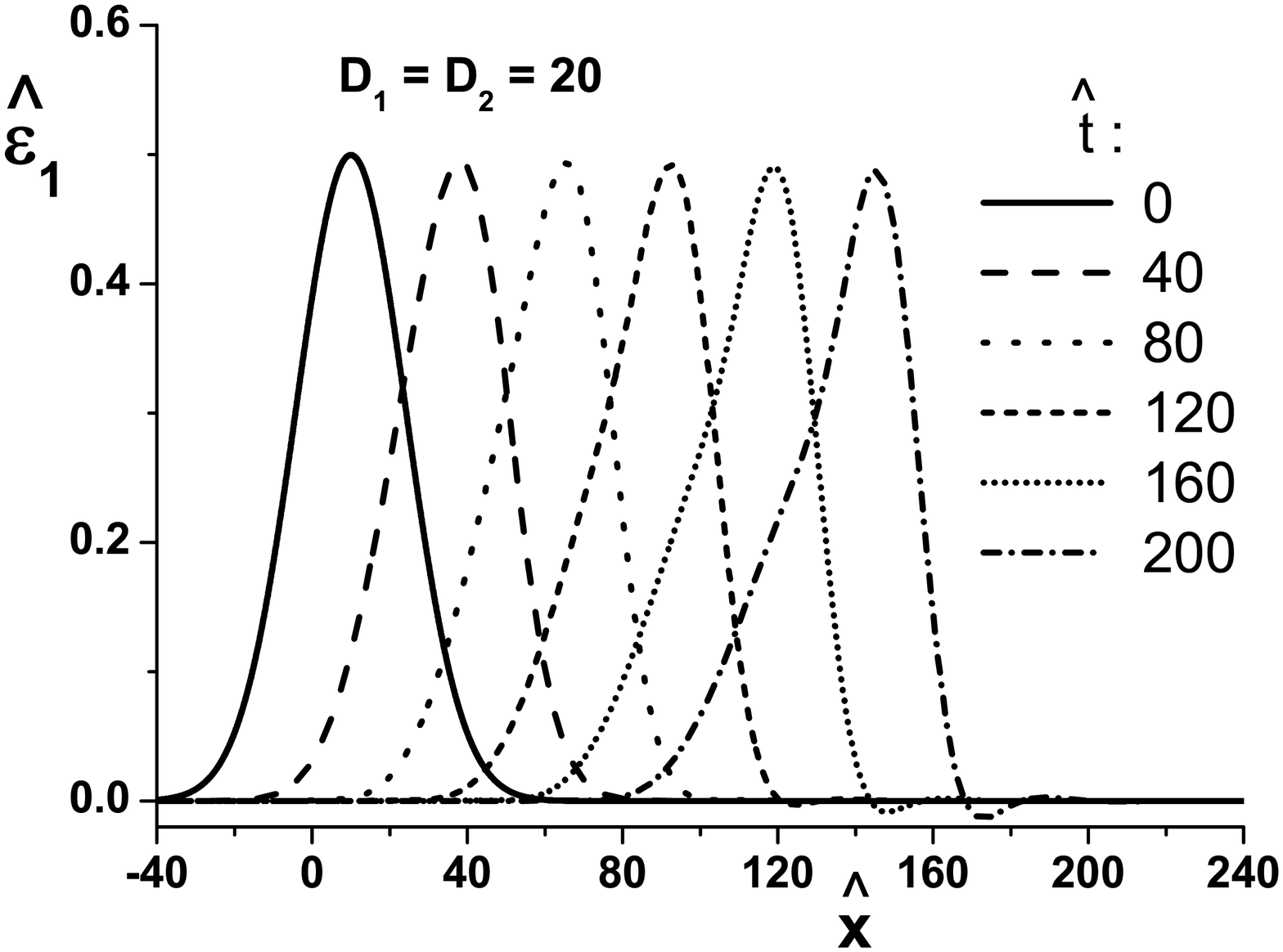}}\\
\subfigure[ ]{\label{fig9c}
\includegraphics[width=0.485\textwidth]{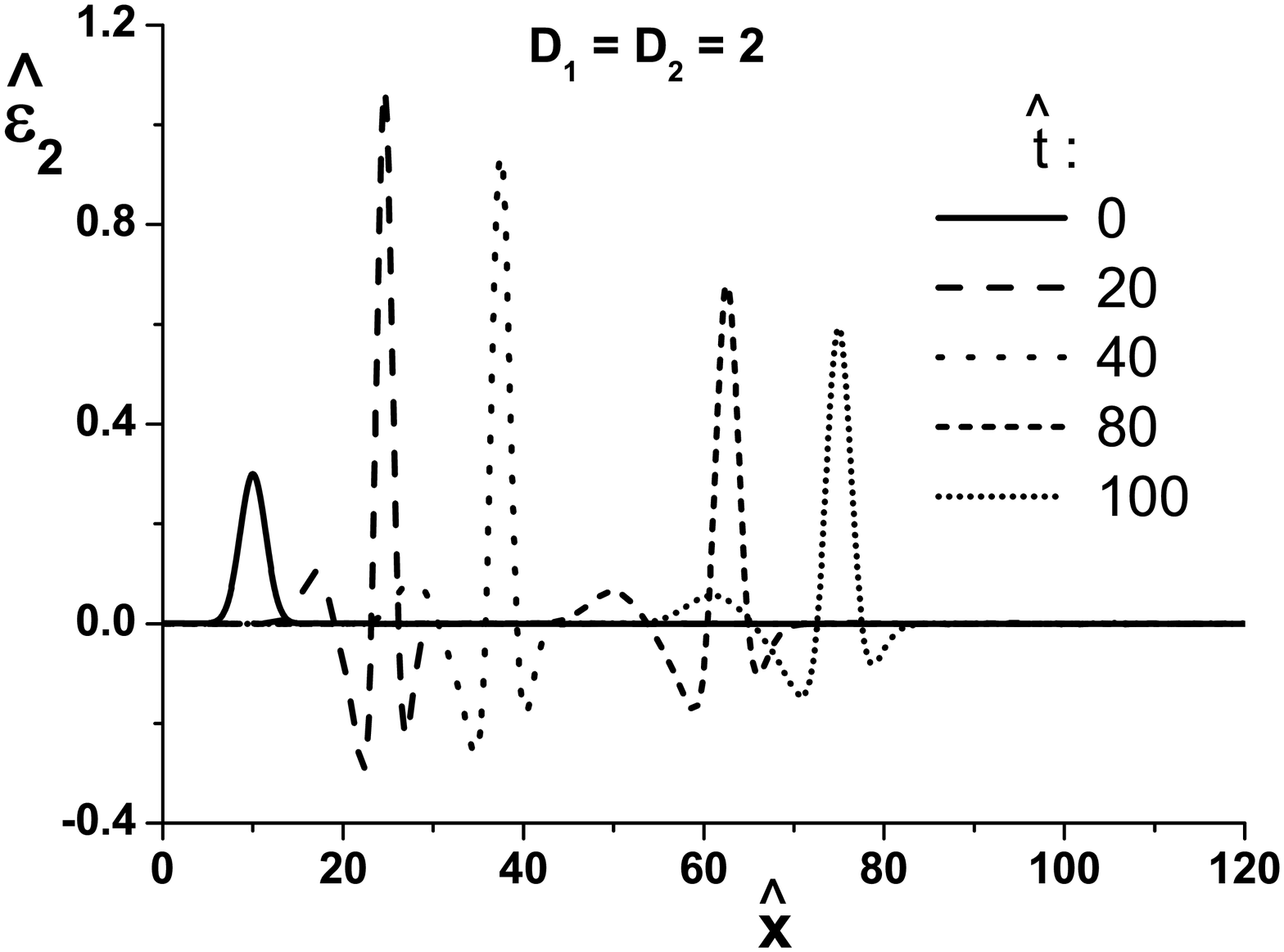}}
\subfigure[ ]{\label{fig9d}
\includegraphics[width=0.485\textwidth]{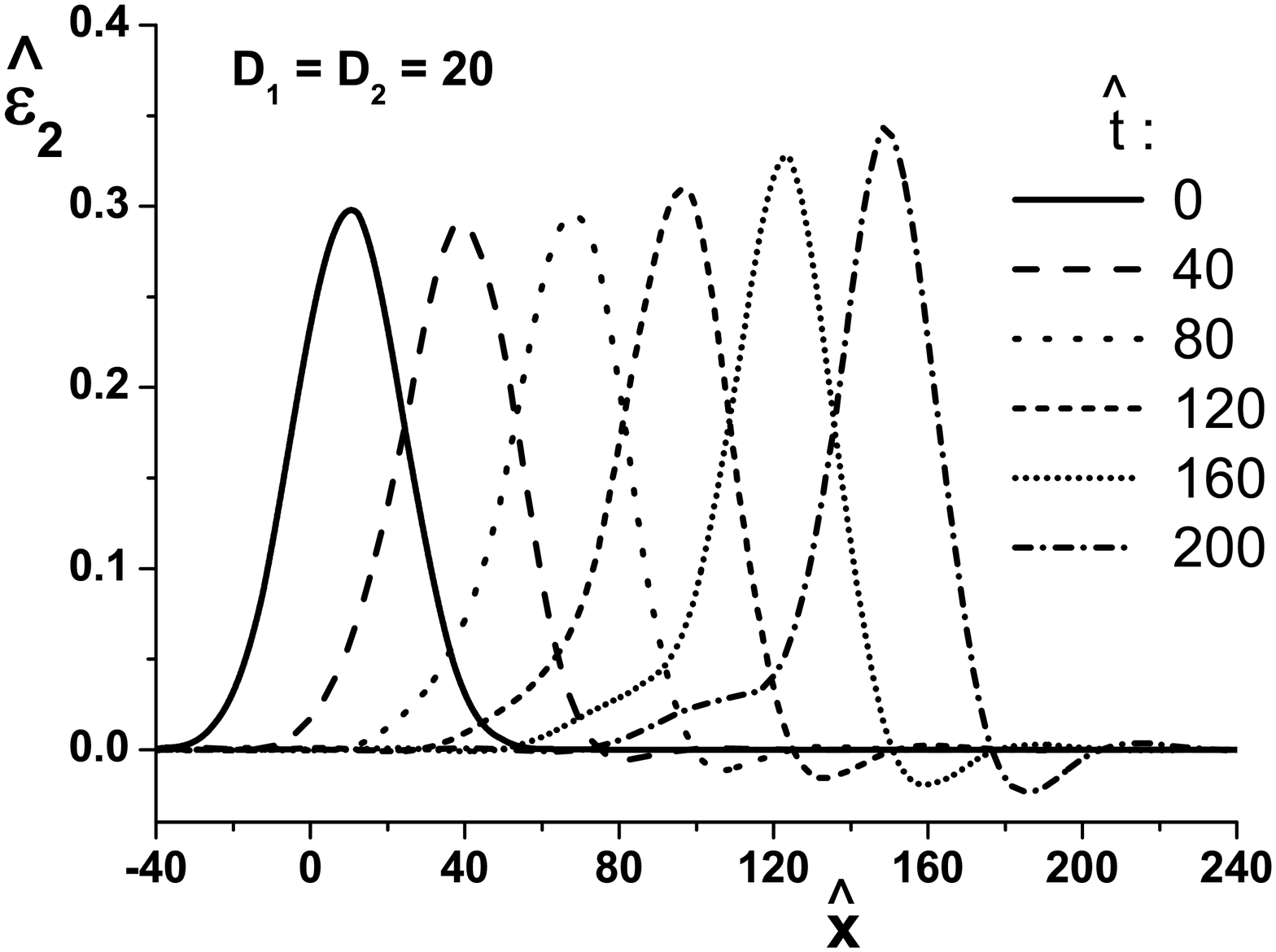}}\\
\subfigure[ ]{\label{fig9e}
\includegraphics[width=0.485\textwidth]{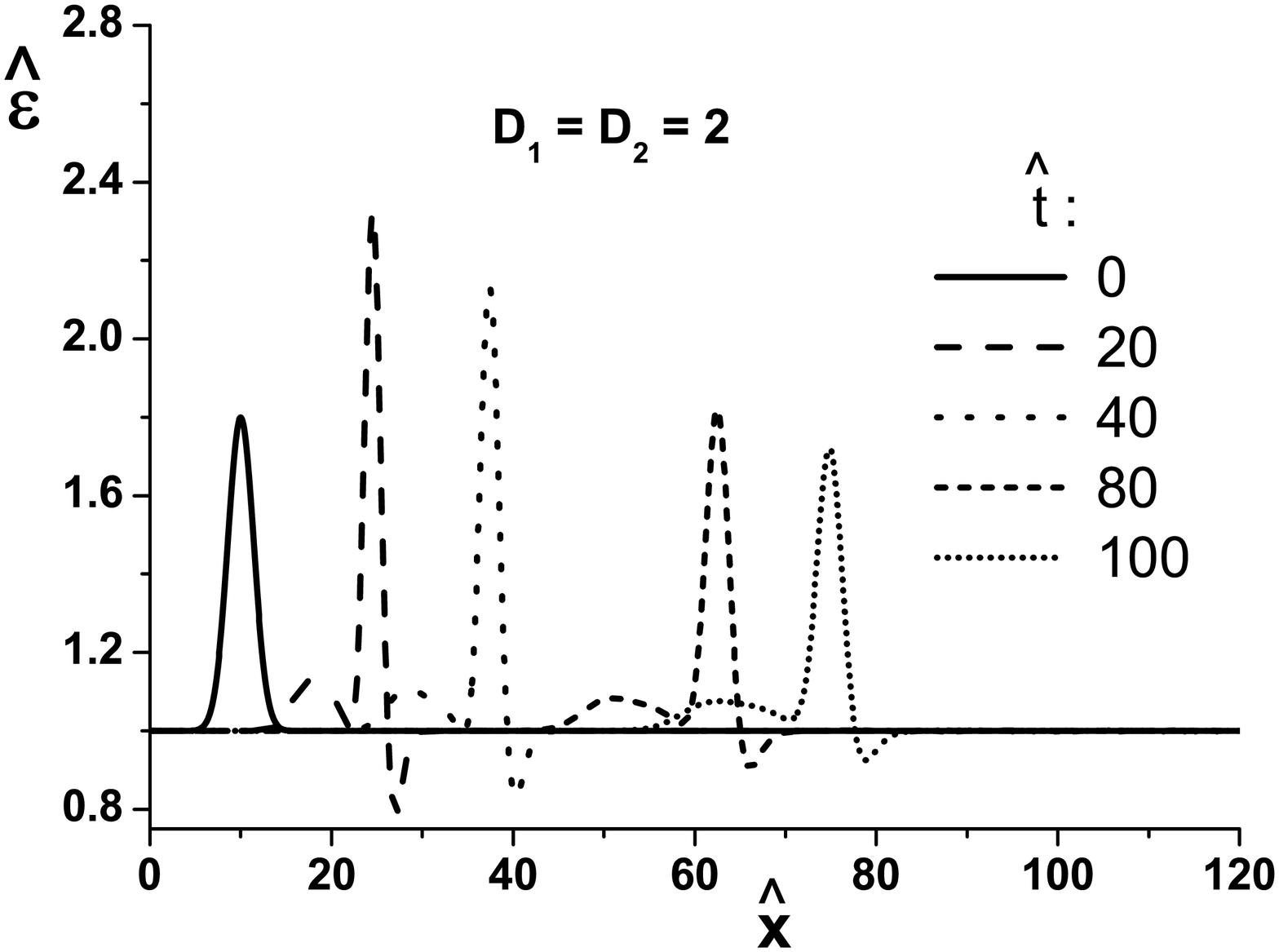}}
\subfigure[ ]{\label{fig9f}
\includegraphics[width=0.485\textwidth]{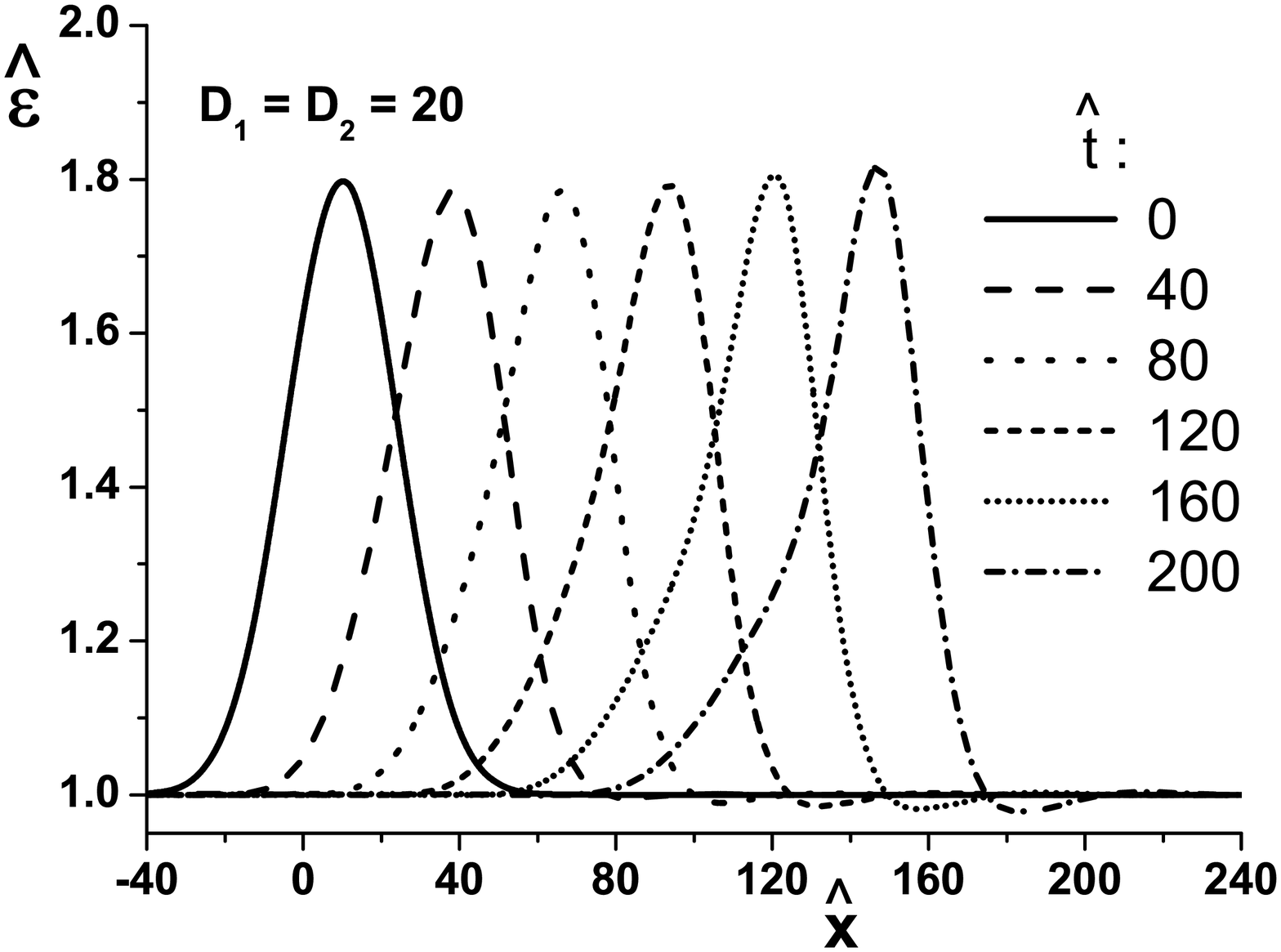}}
\end{center}
\caption{Stability is found by increasing the width of the initial gaussian profiles
(\ref{gle1}) and (\ref{gle2}) with $C_{1}=0.5$ and $C_{2}=0.3$, even for large values of the relaxation time. The plots are the numerical solutions in the nonlinear regime in Eqs.\ (29) (\ref{fig9a} and \ref{fig9b}), and (30) (\ref{fig9c} and \ref{fig9d}) for $\eta_0/s_{0}=1/(4\pi)$ and $\hat{\tau}_{\pi}=200\eta_0/s_{0}$. In \ref{fig9e} and \ref{fig9f}: the complete perturbation. The perturbations survive despite the dissipative effects.}
\label{fig9}
\end{figure}

In Fig. \ref{fig9} we repeat the same calculation for Fig. \ref{fig8}, but considering
a larger value for the relaxation time $\hat{\tau}_{\pi}=200 \eta_{0}/s_{0}$. In Fig. \ref{fig9b} with increasing width the solution displays a soliton-like behavior when compared to Fig. \ref{fig9a}. In Fig. \ref{fig9c} and Fig. \ref{fig9e} we show the case of small width and instabilities in the propagation of the pulse are found. However, we clearly observe in Fig.\ \ref{fig9d} and Fig. \ref{fig9f} that by increasing the width (or, equivalently, by decreasing the initial spatial gradient) one can find a stable propagating pulse even for a large value of the relaxation time. We conclude that even for large values of the relaxation time one can still find a stable nonlinear propagation of the initial gaussian profile, if the initial gradients are sufficiently small, i.e., if the initial gaussian width is small enough. Therefore, in the hydrodynamic limit we find soliton-like solutions of the nonlinear wave equations in Israel-Stewart theory.

In all figures we notice that the numerical solutions of (\ref{burgXTcad}) and (\ref{weeps2onlyad}) do not diverge for long times, i.e., they are not unstable.  The nontrivial study of causality and stability for nonlinear wave equations cannot be performed as simply as it was done in the linear case. Such study is in progress.

\section{Conclusions}

We derived a system of coupled differential equations which describes nonlinear wave perturbations in the energy density of 2nd order conformal fluids.  Our semi-analytical treatment provides a simple (yet nontrivial) picture of how the relaxation time coefficient affects the propagation of sound waves perhaps in a more transparent way than in a complex numerical hydrodynamical simulation.

Our system of differential equations can be easily solved numerically and in certain conditions gives ``soliton-like'' behavior for the initial wave packet evolution.  Our work may be relevant for the understanding of nonlinear perturbations in viscous relativistic hydrodynamics. For instance,
our study of the  deep ``Israel-Stewart limit'' where $\hat{\tau}_{\pi}=120 \eta_0/s_0$ in Fig.\ \ref{fig5} and $\hat{\tau}_{\pi}=200 \eta_0/s_0$ in Fig.\ \ref{fig9} suggest the existence of
an upper bound for $\hat{\tau}_{\pi}$ (for a given $\eta_0/s_{0}$), which marks the onset of a possible instability in the solutions in this case that involves moderately large initial spatial gradients. While the linearized study of wave propagation shows that $\hat{\tau}_\pi$ cannot be much smaller than $\eta_0/s_0$ (due to instabilities), our nonlinear treatment of the wave equation for the energy density in hydrodynamics indicates that in a consistent microscopic theory $\hat{\tau}_\pi$ and $\eta_0/s_0$ must be of comparable magnitude (this is valid, for instance, in the case of kinetic theory calculations). However, we remark that in the ``rigorous" hydrodynamical limit of small spatial gradients, when considering initial gaussian profiles with large widths, it is possible to avoid instabilities in wave propagation, as observed in Fig. \ref{fig9}, while still maintaining the soliton-like solution. Therefore, our nonlinear study suggests that in the case of small spatial gradients, Israel-Stewart theory should support soliton-like wave phenomena.

For most of our investigations we found that the influence of $\hat{\tau}_\pi$ did not determine the overall behavior of wave propagation in the nonlinear regime.  This conclusion agrees with previous investigations in the literature on the small effect of second order transport coefficients in heavy ion collisions \cite{Luzum:2008cw}.  Our only exceptions were the ones that implied unphysical values of relaxation time and large initial spatial gradients. This statement suggests that in physical systems under conditions that are consistent with the hydrodynamic behavior (small gradients), $\hat{\tau}_\pi/(\eta/s)$ should be of order 1 and the effect of the relaxation time on nonlinear wave propagation can be taken to be a small correction.

The differential equations (\ref{burgXTcad}) and (\ref{weeps2onlyad}) are nontrivial alternative approaches to investigate the nonlinear regime of wave propagation in 2nd order conformal hydrodynamics in the Israel-Stewart approximation. However, they are still simple enough to be investigated with simple numerical routines.  For this particular type of study, these equations offer a simple (though clearly limited) alternative to the full numerical hydrodynamical equations. We hope that our work can be used both as a motivation for the search for soliton waves in the full Israel-Stewart equations as well as a possible check of precision of numerical hydrodynamic codes, similar to the analytical solutions found in \cite{hugo,Hatta:2014gqa,Hatta:2014gga}.

It would be interesting to generalize the analysis performed here to include effects from bulk viscosity (i.e., by dropping the underlying conformal invariance of the equations) and different equations of state. Moreover, even though the nonlinear terms in 2nd order hydrodynamics do not contribute to the linearized study, they may play an interesting role in the investigation of nonlinear wave propagation in the QGP but we leave this investigation to a future study.

\begin{acknowledgments}
This work was  partially financed by the Brazilian funding agencies CAPES, CNPq and FAPESP. J.~N.~thanks the Physics Department of the Ohio State University for its hospitality during the time this work was being finished.
\end{acknowledgments}

\section*{Appendix}

Here we focus on the detailed calculations involved in section $B$, subsection $3$.  The RPM method described in section $B$, subsection $1$ is used
in the simplest conformal relativistic Israel-Stewart hydrodynamics equations (\ref{f1}), (\ref{f2}) and (\ref{f3}).

From the property of the dissipative tensor $u_{\mu}\pi^{\mu\nu}=0$  we can write
\begin{equation}
\pi^{tt}=v^{2}\pi^{xx}
\hspace{2cm} \textrm{and} \hspace{2cm}
\pi^{tx}=\pi^{xt}=v\pi^{xx} \, .
\label{constraints}
\end{equation}

Notice that even though the RPM perturbative scheme does take into consideration nonlinearities, we can the general form of the flow, $u_{\mu}  = (\gamma, -\gamma v , 0, 0)$, to write all the possible contributions of the shear tensor components. Since it is traceless, the diagonal components are related as
\begin{eqnarray}
g_{\mu \nu} \pi^{\mu \nu} = \pi^{t t} - \pi^{x x} - \pi^{y y} - \pi^{z z} = 0  \, , \nonumber \\
\pi^{\perp} \equiv \pi^{y y} = \pi^{z z} \, , \nonumber \\
\pi^{\perp} = \frac{(v^{2} - 1 )}{2} \pi^{x x}   \, .
\end{eqnarray}
Using the conservation of energy, the term that contains the dissipative tensor in Eq.\ (\ref{f1}) can be written as a function of the $x$ coordinate (using the previous relations)
\begin{equation}
\pi^{\mu \nu} \sigma_{\mu \nu} = \left(\frac{3}{2} - v^{2} + \frac{3}{2} v^{4}\right) \pi^{x x} \sigma_{x x}  \, .
\end{equation}
The same can be done for the dissipative contribution in the momentum equation Eq. (\ref{f2}):
\begin{equation}
\Delta^{x}_{\mu} \partial_{\nu} \pi^{\mu \nu}  = v \partial_{t} \pi^{x x} + \partial_{x} \pi^{x x}  \, .
\end{equation}
Now, for more general flow patterns the relaxation equation Eq.\ (\ref{f3}) will couple the different components of the dissipative tensor. However, in our particular case regarding this $1+1$ flow pattern, different components do not couple and the relevant term simply becomes
\begin{equation}
\Delta^{x}_{\alpha} \Delta^{x}_{\beta} D \pi^{\alpha \beta} = \gamma^{4}(1 - v^{2})^{2} D \pi^{x x} = D \pi^{x x}  \, .
\end{equation}
Therefore, our analysis is consistent (and simple) and does not require any further approximation regarding the mixing of different shear stress tensor components.

Using $\varepsilon_{0}/ \kappa \, {T_{0}}^{4}=3/4$ and (\ref{constraints}) in (\ref{f1}), (\ref{f2}) and (\ref{f3}), performing the operations $(a)$ to $(c)$, we find:
$$
\sigma \Bigg\lbrace -{\frac{\partial \varepsilon_{1}}{\partial X}}+
{\frac{4}{3}}\,{\frac{\partial v_{1}}{\partial X}} \Bigg\rbrace+
\sigma^{2} \Bigg\lbrace -{\frac{\partial\varepsilon_{2}}{\partial X}}+{\frac{4}{3}}\,{\frac{\partial v_{2}}{\partial X}}
+{\frac{\partial\varepsilon_{1}}{\partial Y}}+v_{1}{\frac{\partial \varepsilon_{1}}{\partial X}}
-{\frac{4}{9}}\,v_{1}{\frac{\partial v_{1}}{\partial X}}
+{\frac{1}{3}} \,\pi^{xx}_{1}{\frac{\partial v_{1}}{\partial X}}\Bigg\rbrace
$$
$$
+ \sigma^{3} \Bigg\lbrace -{\frac{\partial\varepsilon_{3}}{\partial X}}+{\frac{\partial\varepsilon_{2}}{\partial Y}}+v_{1}{\frac{\partial \varepsilon_{2}}{\partial X}}+v_{2}{\frac{\partial \varepsilon_{1}}{\partial X}}-{\frac{4}{9}}\,v_{1}{\frac{\partial v_{2}}{\partial X}}
-{\frac{4}{9}}\,v_{2}{\frac{\partial v_{1}}{\partial X}}
+{\frac{4}{9}}\,v_{1}{\frac{\partial v_{1}}{\partial Y}}+{\frac{4}{3}}\,{\frac{\partial v_{3}}{\partial X}}+{\frac{4}{9}}\,{v_{1}}^{2}{\frac{\partial v_{1}}{\partial X}}
$$
\begin{equation}
-{\frac{1}{9}}\,\pi^{xx}_{1}
{v_{1}}{\frac{\partial v_{1}}{\partial X}}+{\frac{1}{3}}\,\pi^{xx}_{1}{\frac{\partial v_{2}}{\partial X}}
+{\frac{1}{3}}\,\pi^{xx}_{2}{\frac{\partial v_{1}}{\partial X}}\Bigg\rbrace =0 \, ,
\label{f1strexp}
\end{equation}

$$
\sigma \Bigg\lbrace {\frac{\partial\varepsilon_{1}}{\partial X}}-{\frac{4}{3}}\,{\frac{\partial v_{1}}{\partial X}}+{\frac{\partial \pi^{xx}_{1}}{\partial X}}
\Bigg\rbrace+
\sigma^{2} \Bigg\lbrace
-{\frac{4}{3}}\,{\frac{\partial v_{2}}{\partial X}}
+{\frac{\partial \varepsilon_{2}}{\partial X}}
+{\frac{4}{3}}\,{\frac{\partial v_{1}}{\partial Y}}
+{\frac{4}{3}}\,v_{1}{\frac{\partial v_{1}}{\partial X}}-{\frac{1}{3}}\,v_{1}{\frac{\partial \varepsilon_{1}}{\partial X}}
$$
$$
-{\frac{1}{3}}\,
{v_{1}}{\frac{\partial \pi^{xx}_{1}}{\partial X}}-{\frac{1}{3}}\,
\pi^{xx}_{1}{\frac{\partial {v_{1}}}{\partial X}}+{\frac{\partial \pi^{xx}_{2}}{\partial X}} \Bigg\rbrace
+\sigma^{3} \Bigg\lbrace -{\frac{4}{3}}\,{\frac{\partial v_{3}}{\partial X}}+{\frac{4}{3}}\,{\frac{\partial v_{2}}{\partial Y}}+{\frac{4}{3}}\,{v_{1}}{\frac{\partial v_{2}}{\partial X}}+{\frac{4}{3}}\,{v_{2}}{\frac{\partial v_{1}}{\partial X}}-{\frac{4}{9}}\,{v_{1}}^{2}{\frac{\partial v_{1}}{\partial X}}
$$
$$
-{\frac{1}{3}}\,v_{2}{\frac{\partial \varepsilon_{1}}{\partial X}}
-{\frac{1}{3}}\,v_{1}{\frac{\partial \varepsilon_{2}}{\partial X}}
+{\frac{1}{3}}\,v_{1}{\frac{\partial \varepsilon_{1}}{\partial Y}}
+{\frac{\partial\varepsilon_{3}}{\partial X}}+{\frac{\partial \pi^{xx}_{3}}{\partial X}}-{\frac{1}{3}}\,{v_{1}}^{2}{\frac{\partial \pi^{xx}_{1}}{\partial X}}-{\frac{1}{3}}\,{v_{1}}{\frac{\partial \pi^{xx}_{2}}{\partial X}}
$$
\begin{equation}
-{\frac{1}{3}}\,{v_{2}}{\frac{\partial \pi^{xx}_{1}}{\partial X}}+
{\frac{1}{3}}\,{v_{1}}{\frac{\partial \pi^{xx}_{1}}{\partial Y}}
+{\frac{1}{3}}\,{v_{1}}{\frac{\partial \pi^{xx}_{1}}{\partial Y}}
-{\frac{1}{3}}\,\pi^{xx}_{1}{\frac{\partial {v_{2}}}{\partial X}}
-{\frac{1}{3}}\,\pi^{xx}_{1}{v_{1}}{\frac{\partial {v_{1}}}{\partial X}}
-{\frac{1}{3}}\,\pi^{xx}_{2}{\frac{\partial {v_{1}}}{\partial X}}\Bigg\rbrace =0
\label{f2strexp}
\end{equation}
and
$$
\sigma^{1/2} \Bigg\lbrace L\sqrt{3}\,\pi^{xx}_{1}\Bigg\rbrace
+\sigma^{3/2} \Bigg\lbrace -\tilde{\tau}_{\pi}{\frac{\partial \pi^{xx}_{1}}{\partial X}}+
{\frac{4}{3}}{\frac{\tilde{\eta}}{p_{0}}}{\frac{\partial {v_{1}}}{\partial X}}+L\sqrt{3}\,\pi^{xx}_{2}\Bigg\rbrace
$$
$$
+\sigma^{5/2} \Bigg\lbrace -\tilde{\tau}_{\pi}{\frac{\partial \pi^{xx}_{2}}{\partial X}}
+\tilde{\tau}_{\pi}{\frac{\partial \pi^{xx}_{1}}{\partial Y}}
+\tilde{\tau}_{\pi}v_{1}{\frac{\partial \pi^{xx}_{1}}{\partial X}}
+{\frac{4}{3}}\tilde{\tau}_{\pi}\,\pi^{xx}_{1}{\frac{\partial {v_{1}}}{\partial X}}
-{\frac{4}{9}}{\frac{\tilde{\eta}}{p_{0}}}\,v_{1}{\frac{\partial {v_{1}}}{\partial X}}
$$
\begin{equation}
+{\frac{4}{3}}{\frac{\tilde{\eta}}{p_{0}}}\,{\frac{\partial {v_{2}}}{\partial X}}+L\sqrt{3}\,\pi^{xx}_{3}
-{\frac{L}{\sqrt{3}}}\,\pi^{xx}_{1}v_{1}
\Bigg\rbrace =0
\label{f3strexp}
\end{equation}
respectively.  The pressure ${p_{0}}$ is the background pressure.  As described in the step $(d)$, from the $\mathcal{O}({\sigma^{1/2}})$ term in (\ref{f3strexp}) we have:
\begin{equation}
\pi^{xx}_{1} = 0  \,\, .
\label{pi1zero}
\end{equation}
From $\mathcal{O}({\sigma})$ terms of (\ref{f1strexp}) and (\ref{f2strexp}) ( and using (\ref{pi1zero}) ) we find:
\begin{equation}
v_{1}={\frac{3}{4}}\,\varepsilon_{1}  \,\, .
\label{sigma1}
\end{equation}
Applying (\ref{pi1zero}) and (\ref{sigma1}) to the $\mathcal{O}({\sigma^{3/2}})$ term of (\ref{f3strexp}), we find:
\begin{equation}
\pi^{xx}_{2} = -{\frac{\tilde{\eta}}{p_{0}}}
{\frac{1}{L\sqrt{3}}}\,{\frac{\partial {\varepsilon_{1}}}{\partial X}}  \,\, .
\label{sigma3/2}
\end{equation}
Similarly, applying (\ref{pi1zero}), (\ref{sigma1}) and (\ref{sigma3/2}) to the
$\mathcal{O}({\sigma^{2}})$ terms of (\ref{f1strexp}) and (\ref{f2strexp}) we find respectively:
\begin{equation}
{\frac{4}{3}}\,{\frac{\partial v_{2}}{\partial X}}-{\frac{\partial\varepsilon_{2}}{\partial X}}=-{\frac{\partial\varepsilon_{1}}{\partial Y}}
-{\frac{1}{2}}\,\varepsilon_{1}{\frac{\partial\varepsilon_{1}}{\partial X}}
\label{sigma2a}
\end{equation}
and
\begin{equation}
{\frac{4}{3}}\,{\frac{\partial v_{2}}{\partial X}}-{\frac{\partial\varepsilon_{2}}{\partial X}}={\frac{\partial\varepsilon_{1}}{\partial Y}}
+{\frac{1}{2}}\,\varepsilon_{1}{\frac{\partial\varepsilon_{1}}{\partial X}}
-{\frac{\tilde{\eta}\sqrt{3}}{3L p_{0}}}\,{\frac{\partial^{2} {\varepsilon_{1}}}{\partial X^{2}}}
\,\, .
\label{sigma2b}
\end{equation}

Inserting (\ref{pi1zero}), (\ref{sigma1}) and (\ref{sigma3/2}) into the
$\mathcal{O}({\sigma^{5/2}})$ term of (\ref{f3strexp}) we obtain:
\begin{equation}
{\frac{4}{3}}{\frac{\tilde{\eta}}{p_{0}}}\,{\frac{\partial {v_{2}}}{\partial X}}+L\sqrt{3}\,\pi^{xx}_{3}-
{\frac{\tilde{\eta}}{4 p_{0}}}\,\varepsilon_{1}
{\frac{\partial {\varepsilon_{1}}}{\partial X}}+
\tilde{\tau}_{\pi}{\frac{\tilde{\eta}}{L\sqrt{3} \,p_{0}}}\,{\frac{\partial^{2} {\varepsilon_{1}}}{\partial X^{2}}}=0 \,\, .
\label{sigma5/2}
\end{equation}

Equating (\ref{sigma2a}) with (\ref{sigma2b}) we find the following Burgers' equation for
$\varepsilon_{1}(X,T)$:
\begin{equation}
{\frac{\partial\varepsilon_{1}}{\partial Y}}
+{\frac{1}{2}}\,\varepsilon_{1}{\frac{\partial\varepsilon_{1}}{\partial X}}
={\frac{\tilde{\eta}\sqrt{3}}{6L p_{0}}}\,{\frac{\partial^{2} {\varepsilon_{1}}}{\partial X^{2}}}
\,\, .
\label{burgXT}
\end{equation}

The $\mathcal{O}({\sigma^{3}})$ terms of (\ref{f1strexp}) and (\ref{f2strexp}) provide, after using (\ref{pi1zero}), (\ref{sigma1}) and (\ref{sigma3/2}) the following results:
$$
{\frac{4}{3}}\,{\frac{\partial v_{3}}{\partial X}}-{\frac{\partial\varepsilon_{3}}{\partial X}}+{\frac{\partial\varepsilon_{2}}{\partial Y}}
+{\frac{3}{4}}\,\varepsilon_{1}{\frac{\partial\varepsilon_{2}}{\partial X}}
+{\frac{2}{3}}\,v_{2}{\frac{\partial\varepsilon_{1}}{\partial X}}
-{\frac{1}{3}}\,\varepsilon_{1}{\frac{\partial v_{2}}{\partial X}}
$$
\begin{equation}
+{\frac{1}{4}}\,
\varepsilon_{1}{\frac{\partial\varepsilon_{1}}{\partial Y}}
+{\frac{3}{16}}\,
{\varepsilon_{1}}^{2}{\frac{\partial\varepsilon_{1}}{\partial X}}
-{\frac{\tilde{\eta}}{4Lp_{0}\sqrt{3}}}\,
\Bigg({\frac{\partial\varepsilon_{1}}{\partial X}}\Bigg)^{2}=0
\label{sigma3a}
\end{equation}
and
$$
{\frac{\partial\varepsilon_{3}}{\partial X}}-{\frac{4}{3}}\,{\frac{\partial v_{3}}{\partial X}}+{\frac{4}{3}}\,{\frac{\partial v_{2}}{\partial Y}}
-{\frac{1}{4}}\,\varepsilon_{1}{\frac{\partial\varepsilon_{2}}{\partial X}}
+{\frac{2}{3}}\,v_{2}{\frac{\partial\varepsilon_{1}}{\partial X}}
+\varepsilon_{1}{\frac{\partial v_{2}}{\partial X}}
+{\frac{1}{4}}\,\varepsilon_{1}{\frac{\partial\varepsilon_{1}}{\partial Y}}
-{\frac{3}{16}}\,
{\varepsilon_{1}}^{2}{\frac{\partial\varepsilon_{1}}{\partial X}}
$$
\begin{equation}
+{\frac{\tilde{\eta}}{4Lp_{0}\sqrt{3}}}\,
\Bigg({\frac{\partial\varepsilon_{1}}{\partial X}}\Bigg)^{2}
+{\frac{\partial \pi^{xx}_{3}}{\partial X}}+{\frac{\tilde{\eta}}{4Lp_{0}\sqrt{3}}}\,
\varepsilon_{1}{\frac{\partial^{2}\varepsilon_{1}}{\partial X^{2}}}=0 \,\, .
\label{sigma3b}
\end{equation}
Isolating $\pi^{xx}_{3}$ in (\ref{sigma5/2}) and $\partial\varepsilon_{3} / \partial X$ in (\ref{sigma3a}), and then substituting these two results into (\ref{sigma3b}) we obtain
the following equation for $\varepsilon_{2}(X,T)$ and $v_{2}(X,T)$ (considering $\varepsilon_{1}$ previously known from
(\ref{burgXT}) ) :
$$
{\frac{\partial\varepsilon_{2}}{\partial Y}}+
{\frac{4}{3}}\,{\frac{\partial v_{2}}{\partial Y}}
+{\frac{1}{2}}\,\varepsilon_{1}{\frac{\partial\varepsilon_{2}}{\partial X}}
+{\frac{4}{3}}\,v_{2}{\frac{\partial\varepsilon_{1}}{\partial X}}+
{\frac{2}{3}}\,\varepsilon_{1}{\frac{\partial v_{2}}{\partial X}}
$$
$$
+{\frac{1}{2}}\,
\varepsilon_{1}{\frac{\partial\varepsilon_{1}}{\partial Y}}+{\frac{\tilde{\eta}}{4Lp_{0}\sqrt{3}}}\,
\varepsilon_{1}{\frac{\partial^{2}\varepsilon_{1}}{\partial X^{2}}}
+{\frac{\tilde{\eta}}{4Lp_{0}\sqrt{3}}}\,\Bigg[\Bigg({\frac{\partial\varepsilon_{1}}{\partial X}}\Bigg)^{2}+
\varepsilon_{1}{\frac{\partial^{2}\varepsilon_{1}}{\partial X^{2}}}\Bigg]
$$
\begin{equation}
-\tilde{\tau}_{\pi}
{\frac{\tilde{\eta}}{3L^{2}p_{0}}}\,{\frac{\partial^{3}\varepsilon_{1}}{\partial X^{3}}}
-{\frac{4\sqrt{3}\tilde{\eta}}{9Lp_{0}}}\,{\frac{\partial^{2} v_{2}}{\partial X^{2}}}=0
\,\, .
\label{weeps2ev2}
\end{equation}

We have thus a system of wave equations: (\ref{sigma2a}), (\ref{burgXT}) and (\ref{weeps2ev2}) for the three variables: $\varepsilon_{1}(X,T)$, $\varepsilon_{2}(X,T)$ and $v_{2}(X,T)$.  In order to solve it, we shall return to the Cartesian $(x,t)$ space using the (\ref{xtst}) and (\ref{stv}) as described in the step $(d)$ of the RPM.  So, (\ref{sigma2a}), (\ref{burgXT}) and (\ref{weeps2ev2}) are rewritten as:
\begin{equation}
{\frac{4}{3}}\,{\frac{\partial }{\partial x}} \hat{v}_{2}-{\frac{\partial }{\partial x}}\hat{\varepsilon}_{2}=-\sqrt{3}
{\frac{\partial}{\partial t}}\hat{\varepsilon}_{1}-{\frac{\partial}{\partial x}}\hat{\varepsilon}_{1}
-{\frac{1}{2}}\,\hat{\varepsilon}_{1}{\frac{\partial}{\partial x}}\hat{\varepsilon}_{1}
\,\, ,
\label{sigma2ac}
\end{equation}
\\
\begin{equation}
{\frac{\partial}{\partial t}}\hat{\varepsilon}_{1}+{\frac{1}{\sqrt{3}}}\,{\frac{\partial}{\partial x}}
\hat{\varepsilon}_{1}
+{\frac{1}{2\sqrt{3}}}\,\hat{\varepsilon}_{1}{\frac{\partial}{\partial x}}\hat{\varepsilon}_{1}
={\frac{\eta}{6 p_{0}}}\,{\frac{\partial^{2} }{\partial x^{2}}}\hat{\varepsilon}_{1}
\label{burgXTc}
\end{equation}
\\
and
$$
{\frac{\partial}{\partial t}}\hat{\varepsilon_{2}}+
{\frac{1}{\sqrt{3}}}\,{\frac{\partial}{\partial x}}
\hat{\varepsilon}_{2}+
{\frac{4}{3}}\,{\frac{\partial }{\partial t}}\hat{v}_{2}
+{\frac{4}{3\sqrt{3}}}\,{\frac{\partial }{\partial x}}\hat{v}_{2}
+{\frac{1}{2\sqrt{3}}}\,\hat{\varepsilon}_{1}{\frac{\partial}{\partial x}}\hat{\varepsilon}_{2}
+{\frac{4}{3\sqrt{3}}}\,\hat{v}_{2}{\frac{\partial}{\partial x}}\hat{\varepsilon}_{1}
+{\frac{2}{3\sqrt{3}}}\,\hat{\varepsilon}_{1}{\frac{\partial }{\partial x}}\hat{v}_{2}
$$
$$
+{\frac{1}{2}}\,
\hat{\varepsilon}_{1}{\frac{\partial }{\partial t}}\hat{\varepsilon}_{1}
+{\frac{1}{2\sqrt{3}}}\,
\hat{\varepsilon}_{1}{\frac{\partial }{\partial x}}\hat{\varepsilon}_{1}
+{\frac{\eta}{4p_{0}}}\,\hat{\varepsilon}_{1}{\frac{\partial^{2}}{\partial x^{2}}}\hat{\varepsilon}_{1}
+{\frac{\eta}{4p_{0}}}\,\Bigg[
\Bigg({\frac{\partial }{\partial x}}\hat{\varepsilon}_{1}\Bigg)^{2}+
\hat{\varepsilon}_{1}{\frac{\partial^{2}}{\partial x^{2}}}\hat{\varepsilon}_{1}\Bigg]
$$
\begin{equation}
-\tau_{\pi}{\frac{\eta}{3\sqrt{3} \,p_{0}}}\,{\frac{\partial^{3}}{\partial x^{3}}}\hat{\varepsilon}_{1}
-{\frac{4\eta}{9p_{0}}}\,{\frac{\partial^{2}}{\partial x^{2}}}
\hat{v}_{2}=0 \,\, .
\label{weeps2ev2c}
\end{equation}
The three equations above are for the dimensionless variables $\hat{\varepsilon}_{1} \equiv \sigma\varepsilon_{1}$, $\hat{\varepsilon}_{2} \equiv \sigma^{2} \varepsilon_{2}$ as defined in (\ref{enerdenexp}) and $\hat{v}_{2} \equiv \sigma^{2} v_{2}$ from
(\ref{vexp}).  Inserting (\ref{burgXTc}) into (\ref{sigma2ac}) we obtain:
\begin{equation}
{\frac{\partial}{\partial x}}\hat{v}_{2}={\frac{3}{4}}\,{\frac{\partial}{\partial x}}\hat{\varepsilon}_{2}-{\frac{\eta\sqrt{3}}{8p_{0}}}\,{\frac{\partial^{2} }{\partial x^{2}}}\hat{\varepsilon}_{1}
\label{burgXTcamalg}
\end{equation}
which, considering the constant of integration equals to zero yields the the following relation:
\begin{equation}
\hat{v}_{2}={\frac{3}{4}}\,\hat{\varepsilon}_{2}-{\frac{\eta\sqrt{3}}{8p_{0}}}\,{\frac{\partial }{\partial x}}\hat{\varepsilon}_{1} \,\, .
\label{v2}
\end{equation}
Calculating the spatial derivative of (\ref{burgXTc}) we have:
\begin{equation}
{\frac{\partial}{\partial t}}{\frac{\partial}{\partial x}}\hat{\varepsilon}_{1}=-{\frac{1}{\sqrt{3}}}\,{\frac{\partial^{2}}{\partial x^{2}}}
\hat{\varepsilon}_{1}
-{\frac{1}{2\sqrt{3}}}\,\Bigg[\Bigg({\frac{\partial}{\partial x}}\hat{\varepsilon}_{1}\Bigg)^{2}+\hat{\varepsilon}_{1}{\frac{\partial^{2}}{\partial x^{2}}}\hat{\varepsilon}_{1}\Bigg]
+{\frac{\eta}{6 p_{0}}}\,{\frac{\partial^{3} }{\partial x^{3}}}\hat{\varepsilon}_{1}
\label{burgXTcdx}
\end{equation}

Substituting (\ref{v2}) and (\ref{burgXTcdx}) in (\ref{weeps2ev2c}) we find:
$$
{\frac{\partial}{\partial t}}\hat{\varepsilon_{2}}+
{\frac{1}{\sqrt{3}}}\,{\frac{\partial}{\partial x}}
\hat{\varepsilon}_{2}
+{\frac{1}{2\sqrt{3}}}\,\hat{\varepsilon}_{1}{\frac{\partial}{\partial x}}\hat{\varepsilon}_{2}
-{\frac{\eta}{6 p_{0}}}\,{\frac{\partial^{2} }{\partial x^{2}}}\hat{\varepsilon}_{2}
+{\frac{1}{2\sqrt{3}}}\,\hat{\varepsilon}_{2}{\frac{\partial}{\partial x}}\hat{\varepsilon}_{1}
$$
\begin{equation}
+{\frac{\eta}{12 p_{0}}}\,
\hat{\varepsilon}_{1}{\frac{\partial^{2}}{\partial x^{2}}}\hat{\varepsilon}_{1}
+{\frac{1}{4}}\hat{\varepsilon}_{1}{\frac{\partial}{\partial t}}\hat{\varepsilon}_{1}
+{\frac{1}{4\sqrt{3}}}
\hat{\varepsilon}_{1}{\frac{\partial}{\partial x}}\hat{\varepsilon}_{1}
+{\frac{\eta}{6 p_{0}}}
\Bigg[{\frac{\eta \sqrt{3}}{12 p_{0}}}\,
 -{\frac{\tau_{\pi}}{\sqrt{3}}}\Bigg]{\frac{\partial^{3} }{\partial x^{3}}}\hat{\varepsilon}_{1}=0 \,\, .
\label{weeps2only}
\end{equation}
Finally, the set of equations for the small perturbations in energy density: $\hat{\varepsilon}_{1}$ and $\hat{\varepsilon}_{2}$ as described by (\ref{enerdenexp}),
is given by the Burgers' equation (\ref{burgXTc}) and the equation (\ref{weeps2only}).

Using the dimensionless variables $\hat{x}=x \, T_{0}$, $\hat{t}=t \, T_{0}$, $\hat{\tau}_{\pi}=T_{0}\,\tau_{\pi}$ and recalling to the Gibbs relation $p_{0}=T_{0}s_{0}/4$, we rewrite (\ref{burgXTc}) and (\ref{weeps2only}) as:

\begin{equation}
{\frac{\partial}{\partial \hat{t}}}\hat{\varepsilon}_{1}+{\frac{1}{\sqrt{3}}}\,{\frac{\partial}{\partial \hat{x}}}
\hat{\varepsilon}_{1}
+{\frac{1}{2\sqrt{3}}}\,\hat{\varepsilon}_{1}{\frac{\partial}{\partial \hat{x}}}\hat{\varepsilon}_{1}
={\frac{\chi}{2}}\,{\frac{\partial^{2} }{\partial \hat{x}^{2}}}\hat{\varepsilon}_{1}
\label{burgXTcada}
\end{equation}
and
$$
{\frac{\partial}{\partial \hat{t}}}\hat{\varepsilon_{2}}+
{\frac{1}{\sqrt{3}}}\,{\frac{\partial}{\partial \hat{x}}}
\hat{\varepsilon}_{2}
+{\frac{1}{2\sqrt{3}}}\,\hat{\varepsilon}_{1}{\frac{\partial}{\partial \hat{x}}}\hat{\varepsilon}_{2}
-{\frac{\chi}{2}} \,{\frac{\partial^{2} }{\partial {\hat{x}}^{2}}}\hat{\varepsilon}_{2}
+{\frac{1}{2\sqrt{3}}}\,\hat{\varepsilon}_{2}{\frac{\partial}{\partial \hat{x}}}\hat{\varepsilon}_{1}
+{\frac{\chi}{4}}\,
\hat{\varepsilon}_{1}{\frac{\partial^{2}}{\partial {\hat{x}}^{2}}}\hat{\varepsilon}_{1}
$$
\begin{equation}
+{\frac{1}{4}}\,
\hat{\varepsilon}_{1}{\frac{\partial}{\partial \hat{t}}}\hat{\varepsilon}_{1}
+{\frac{1}{4\sqrt{3}}}\,
\hat{\varepsilon}_{1}{\frac{\partial}{\partial \hat{x}}}\hat{\varepsilon}_{1}
+{\frac{\chi}{2}}
\Bigg[{\frac{\chi \, \sqrt{3}}{4}}
 -{\frac{\hat{\tau}_{\pi}}{\sqrt{3}}}\Bigg]{\frac{\partial^{3} }{\partial {\hat{x}}^{3}}}\hat{\varepsilon}_{1}=0\,.
\label{weeps2onlyada}
\end{equation}

\end{document}